\documentclass[reqno]{amsart}

\RequirePackage{hyperref}

\title[Embeddability into relational lattices]{
  Embeddability into relational lattices \\is
  undecidable}

\author[Luigi Santocanale]{Luigi Santocanale}
\address{Luigi Santocanale\\
LIF, CNRS UMR 7279, Aix-Marseille Universit\'e}
\email{luigi.santocanale@lif.univ-mrs.fr}

\input macros.tex

\begin{document}
\maketitle

\begin{abstract}
The natural join and the inner union operations combine relations of a
database. Tropashko and Spight realized that these two operations are
the meet and join operations in a class of lattices, known by now as
the relational lattices. They proposed then lattice theory as an
algebraic approach to the theory of databases alternative to the
relational algebra.  Litak et al. proposed an axiomatization of
relational lattices over the signature that extends the pure lattice
signature with a constant and argued that the quasiequational theory
of relational lattices over this extended signature is undecidable.

We prove in this paper that embeddability is undecidable for
relational lattices. More precisely, it is undecidable whether a
finite \si lattice can be embedded into a relational lattice. Our
proof is a reduction from the coverability problem of a multimodal
frame by a universal product frame and, indirectly, from the
representability problem for relation algebras.

As corollaries we obtain the following results: the quasiequational
theory of relational lattices over the pure lattice signature is
undecidable and has no finite base; there is a quasiequation over the
pure lattice signature which holds in all the finite relational
lattices but fails in an infinite relational lattice.

\end{abstract}

\section{Introduction}

The natural join and the inner union operations combine relations
(i.e. tables) of a database. Most of today's web programs query their
databases making repeated use of the natural join and of the union, of
which the inner union is a mathematically well behaved variant.
Tropashko and Spight realized
\cite{Tropashko2005,Tropashko2008} that these two operations are the
meet and join operations in a class of lattices, known by now as the
class of relational lattices. They proposed then lattice theory as an
algebraic approach, alternative to Codd's relational algebra
\cite{codd70}, to the theory of databases.

An important first attempt to axiomatize these lattices is due to
Litak, Mikul\'as, and Hidders \cite{LitakMHjlamp}. These authors
propose an axiomatization, comprising equations and quasiequations, in
a signature that extends the pure lattice signature with a constant,
the header constant. A main result of that paper is that the
quasiequational theory of relational lattices is undecidable in this
extended signature. Their proof mimics Maddux's proof that the
equational theory of cylindric algebras of dimension $n \geq 3$ is
undecidable \cite{Maddux1980}.

We have investigated in \cite{San2016} equational axiomatizations for
relational lattices using as tool the duality theory for finite
lattices developed in \cite{San09:duality}. A conceptual contribution
from \cite{San2016} is to make explicit the similarity between the
developing theory of relational lattices and the well established
theory of combination of modal logics, see
e.g. \cite{Kurucz2007}. This was achieved on the syntactic side, but
also on the semantic side, by identifying some key properties of the
structures dual to the finite atomistic lattices in the variety
generated by the relational lattices, see \cite[Theorem~7]{San2016}.
These properties make the dual structures into frames for commutator
multimodal logics in a natural way.

In this paper we exploit this similarity to transfer results from the
theory of multidimensional modal logics to lattice theory.  Our main
result is that \emph{it is undecidable whether a finite subdirectly
  irreducible lattice can be embedded into a relational lattice}. We
prove this statement by reducing to it the coverability problem of a
frame by a universal $\Sfive^{3}$-product frame, a problem shown to be
undecidable in \cite{HHK2002}. As stated there, the coverability
problem is---in light of standard duality theory---a direct
reformulation of the representability problem of finite simple
relation algebras, problem shown to be undecidable by Hirsch and
Hodkinson \cite{HH2001}.

Our main result and its proof allow us to derive further
consequences. Firstly, we refine the undecidability theorem of
\cite{LitakMHjlamp} and prove that \emph{the quasiequational theory of
  relational lattices in the pure lattice signature is undecidable} as
well and \emph{has no finite base}.  Then we argue that \emph{there is
  a quasiequation that holds in all the finite relational lattices,
  but fails in an infinite one}. For the latter result, we rely on the
work by Hirsch, Hodkinson, and Kurucz \cite{HHK2002} who constructed a
finite $3$-multimodal frame which has no finite \pmorphism from a
finite universal $\Sfive^{3}$-product frame, but has a \pmorphism from
an infinite one.
On the methodological side, we wish to point out our use of generalized
ultrametric spaces to tackle these problems. A key idea in the proof
of the main result is the characterization of universal
$\Sfive^{A}$-product frames as \Pc generalized ultrametric spaces with
distance valued in the Boolean algebra $P(A)$, a characterization that
holds when  $A$ is
finite.

\smallskip

The paper is structured as follows. We recall in
Section~\ref{sec:defselconcepts} few definitions and facts on frames
and lattices. Relational lattices are introduced in
Section~\ref{sec:rellattices}.  In Section~\ref{sec:overview} we give
an outline of the proof of our main technical result---the
undecidability of embeddability of a finite \si lattice into a
relational lattice---and derive from it the other results. In
Section~\ref{sec:onedirection} we show how to construct a lattice from
a frame and use functoriality of this construction to argue that such
lattice embeds into a relational lattice whenever the frame is a
\pmorphic image of a universal product frame. The proof of the
converse statement is carried out in Section~\ref{sec:converse}. The
technical tools needed to prove the converse are developed
Sections~\ref{sec:ultrametric} and~\ref{sec:intervals}. The theory of
generalized ultrametric spaces over a powerset Boolean algebra and the
aforementioned characterization of $\Sfive^{A}$-product frames as \Pc
spaces over $P(A)$ appear in Section~\ref{sec:ultrametric}. In
Section~\ref{sec:intervals} we study embeddings of finite \si lattices
into relational lattices and prove that we can assume that these
embeddings preserve bounds.  This task is needed so to exclude the
constants $\bot$ and $\top$ (denoting the bounds) from the signature
of lattice theory.

\section{Frames and lattices}
\label{sec:defselconcepts}

\paragraph{\bf Frames.}
Let $A$ be a set of actions.  An \emph{$A$-multimodal frame} (briefly,
an $A$-\emph{frame} or a \emph{frame}) is a structure
$\FF = \langle X_{\FF}, \set{R_{a} \mid a \in A}\rangle$ where, for
each $a \in A$, $R_{a}$ is a binary relation on $X_{\FF}$. We say that
an $A$-frame is \emph{\Sfour} if each $R_{a}$ is reflexive and transitive.
If $\FF_{0}$ and $\FF_{1}$ are two $A$-frames, then a
\emph{\pmorphism} from $\FF_{0}$ to $\FF_{1}$ is a function
$\psi : X_{\FF_{0}}\rto X_{\FF_1}$ such that, for each $a \in A$,
\begin{itemize}
\item if $xR_{a} y$, then $\psi(x)R_{a}\psi(y)$,
\item if $\psi(x)R_{a}z$, then $xR_{a}y$ for some $y$ with
  $\psi(y) = z$.
\end{itemize}
Let us mention that $A$-multimodal frames and \pmorphism{s} form a
category.

\medskip

A frame $\FF$ is said to be \emph{\rooted} (or \emph{initial}, see
\cite{Sambin99}) if there is $f_{0} \in X_{\FF}$ such that every other
$f \in X_{\FF}$ is reachable from $f_{0}$. We say that an $A$-frame
$\FF$ is \emph{full} if, for each $a \in A$, there exists
$f,g \in X_{\FF}$ such that $f \neq g$ and $f R_{a} g$.
If $G = (V,D)$ is a directed graph, then we shall say that $G$ is
\rooted if it is \rooted as a unimodal frame.

A particular class of frames we shall deal with are the
\emph{universal $\Sfive^{A}$-product frames}. These are the frames
$\U$ with $X_{\U} = \prod_{a \in A} X_{a}$ and $x R_{a} y$ if and
only if $x_{i} = y_{i}$ for each $i\neq a$, where
$x := \langle x_{i}\mid i \in A \rangle$ and
$y:= \langle y_{i}\mid i \in A \rangle$.

Let $\alpha \subseteq A$, $\FF$ be an $A$-frame, $x, y \in \X$. An
\emph{$\alpha$-path} from $x$ to $y$ is a sequence
$x = x_{0}R_{a_{0}}x_{1}\ldots x_{k-1}R_{a_{k-1}} x_{k} = y$ with
$\set{a_{0},\ldots ,a_{k-1}} \subseteq \alpha$. We use then the
notation $x \xrightarrow{\alpha} y$ to mean that there is an
$\alpha$-path from $x$ to $y$. Notice that if $\FF$ is an \Sfour
$A$-frame, then $x \xrightarrow{\set{a}} y$ if and only if
$x R_{a} y$.

\bigskip

\paragraph{\bf Orders and lattices.}
We assume some basic knowledge of order and lattice theory as
presented in standard monographs \cite{DP02,GLT2}. Most of the tools
we use in this paper originate from the monograph \cite{FJN} and have
been further developed in \cite{San09:duality}.

\smallskip

A \emph{lattice} is a poset $L$ such that every finite non-empty
subset $X \subseteq L$ admits a smallest upper bound $\bv X$ and a
greatest lower bound $\bigwedge X$. 
A lattice can also be understood as a
structure $\A$ for the functional signature $(\vee,\land)$, such that
the interpretations of these two binary function symbols both give
$\A$ the structure of an idempotent commutative semigroup, the two
semigroup structures being connected by the absorption laws
$x \land (y \vee x) = x$ and $x \vee (y \land x) = x$.
Once a lattice is presented as such structure, the order is recovered
by stating that $x \leq y$ holds if and only if $x \land y= x$.

A lattice $L$ is \emph{complete} if any subset $X \subseteq L$ admits
a smallest upper bound $\bv X$. It can be shown that this condition
implies that any subset $X \subseteq L$ admits a greatest lower bound
$\bigwedge X$. A lattice is \emph{bounded} if it has a least element
$\bot$ and a greatest element $\top$. A complete lattice (in
particular, a finite lattice) is bounded, since $\bigvee \emptyset$
and $\bigwedge \emptyset$ are, respectively, the least and greatest
elements of the lattice.

If $P$ and $Q$ are partially ordered sets, then a function
$f : P \rto Q$ is \emph{\op} (or \emph{monotone}) if $p \leq p'$
implies $f(p) \leq f(p')$.  If $L$ and $M$ are lattices, then a
function $f : L \rto M$ is a \emph{lattice morphism} if it preserves
the lattice operations $\vee$ and $\land$. A lattice morphism is
always \op. A lattice morphism $f : L \rto M$ between bounded lattices
$L$ and $M$ is \emph{\bp} if $f(\bot) = \bot$ and $f(\top) = \top$.
A function $g : Q \rto P$ is said to be \emph{\la} to an \op
$f : P \rto Q$ if $g(q) \leq p$ holds if and only if $q \leq f(p)$
holds; such a \la, when it exists, is unique.  If $L$ is finite, $M$
is bounded, and $f : L \rto M$ is a \bp lattice morphism, then a \la
to $f$ always exists and preserves the constant $\bot $ and the
operation $\vee$.

A \emph{Moore family on a set $U$} is a collection ${\cal F}$ of
subsets of $U$ which is closed under arbitrary intersections.  Given a
Moore family ${\cal F}$ on $U$, the correspondence sending
$Z \subseteq U$ to
$\closure{Z}:= \bigcap \set{Y \in {\cal F} \mid Z \subseteq Y }$ is a
closure operator on $U$, that is, an \op inflationary and idempotent
endofunction of $P(U)$.  The subsets in ${\cal F}$, called the
\emph{closed sets}, are exactly the fixpoints of this closure
operator.  We can give ${\cal F}$ a lattice structure by defining
\begin{align}
  \label{eq:opsMooreFamily}
  \bigwedge
  \, X & := \bigcap X\,, &
  \bigvee
  \,X & := \closure{\bigcup X}\,.
\end{align}

Let $L$ be a complete lattice. An element $j \in L$ is 
\emph{\cjirr} if $j = \bv X$ implies $j \in X$, for each
$X \subseteq L$; the set of \cjirr elements of $L$ is denoted here
$\Ji(L)$. A complete lattice is \emph{spatial} if every element is the
join of the \cjirr elements below it.  An element $j \in \Ji(L)$ is
said to be \emph{\jp} if $j \leq \bv X$ implies $j \leq x$ for some
$x \in X$, for each finite subset $X$ of $L$. If $x$ is not \jp, then
we say that $x$ is \emph{\njp}.
An \emph{atom} of a
lattice $L$ is an element of $L$ such that $\bot$ is the only element
strictly below it.
A spatial lattice is \emph{atomistic} if every
element of $\Ji(L)$ is an atom.

For $j \in \Ji(L)$, a \emph{join-cover} of $j$ is a subset
$X \subseteq L$ such that $j \leq \bv X$. For $X, Y \subseteq L$, we
say that $X$ \emph{refines} $Y$, and write $X \refines Y$, if for all
$x \in X$ there exists $y \in Y$ such that $x \leq y$. A join-cover
$X$ of $j$ is said to be \emph{minimal} if $j \leq \bv Y$ and
$Y \refines X$ implies $X \subseteq Y$; we write $j \mcovered X$ if
$X$ is a \mjc of $j$. In a spatial lattice, if $j \mcovered X$, then
$X \subseteq \Ji(L)$. If $j \mcovered X$, then we say that $X$ is a
\emph{non-trivial} \mjc of $j$ if $X\neq \set{j}$. 
The word perfect is used in the lattice-theoretic literature with
different meanings \cite{dunn2005,Adaricheva2011}.
We use here something different:
\begin{definition}
  We say that a complete lattice is \emph{\pperfect} if it is spatial
  and for each $j \in \Ji(L)$ and $X \subseteq L$, if $j \leq \bv X$,
  then $Y \refines X$ for some $Y$ such that $j \mcovered Y$.  The
  \emph{OD-graph} of a \pperfect lattice $L$ is the structure
  $\langle \Ji(L),\leq,\mcovered \rangle$.
\end{definition}
That is, in a \pperfect lattice every cover refines to a minimal
one.
With respect to analogous definitions, such as that of a lattice with
the \emph{$\Sigma$-\mjc refinement property} \cite{wehrung:MCRP} or
that of a strongly spatial lattice \cite{SW2013}, we do
not require here that the set $Y$ in the relation $j \mcovered Y$ is
finite, nor that, for a given $j$, there are a finite number of these
sets.
    
Notice that every finite lattice is \pperfect.  If $L$ is a \pperfect
lattice, then we say that $X \subseteq \Ji(L)$ is \emph{closed} if it
is a downset and $j \mcovered C \subseteq X$ implies $j \in X$.
Closed subsets of $\Ji(L)$ form a Moore family.
The interest of considering \pperfect lattices stems from the
following representation theorem stated in \cite{Nation90} for finite
lattices; its generalization to \pperfect lattices is straightforward.
\begin{theorem}
  \label{thm:nation}
  Let $L$ be a
  \pperfect lattice and 
  let $\L(\Ji(L),\leq,\mcovered)$ be the lattice of
  closed subsets of $\Ji(L)$. The mapping $l \mapsto \set{j \in
    \Ji(L) \mid j \leq l}$ is a lattice isomorphism from $L$ to
  $\L(\Ji(L),\leq,\mcovered)$.
\end{theorem}
\begin{proof}
  Let $f(l) := \set{j \in \Ji(L) \mid j \leq l}$. Clearly $f(l)$ is a
  downset, let us verify that it is closed as well: if  $j \mcovered C \subseteq f(l)$, then $C
  \refines l$ and $j \leq \bv C \leq l$, so $j \in f(l)$.

  Observe now that $f$ is \op; to see that $f$ is an order
  isomorphism we argue that $\bv f(l) = l$ and $f(\bv X) = X$, when
  $X$ is closed subset of $\Ji(L)$. 

  If $j \leq \bv f(l)$, then $j \mcovered C \refines f(l)$; since
  $f(l)$ is a downset, $C \subseteq f(l)$ follows and therefore
  $j \in f(l)$, since $f(l)$ is closed; that is, we have $j \leq l$.
  By spatiality, we have therefore that $\bv f(l) \leq l$; equality
  follows since clearly $l \leq \bv f(l)$.  For the second relation,
  if $j \in X$, then $j \leq \bv X$ and $j \in f(\bv X)$, so
  $X \subseteq f(\bv X)$.  Conversely, if $j \in f(\bv X)$, then
  $j \leq \bv X$ and $j \mcovered C \refines X$. Since $X$ is a
  downset, then $C \subseteq X$ and since $X$ is closed, then
  $j \in X$. Thus $f(\bv X) \subseteq X$ and equality holds.
\end{proof}

It was shown in \cite{San09:duality} how to extend this representation
theorem to a duality between the category of finite lattices and the
category of OD-graphs. We develop next some observations about
\pperfect lattices, that generalize well known facts on finite
lattices. We shall need these observations mainly in the course of
Section~\ref{sec:intervals}.

For a lattice $L$, a \emph{principal ideal} of $L$ is a subset of the
form $\downset{\,l} := \set{x \in L \mid x \leq l}$.
\begin{lemma}
  \label{lemma:pipperfect}
  If $L$ is a \pperfect lattice, then every principal ideal
  $\downset{\,l}$, $l \in L$, is also \pperfect. We have
  $\Ji(\downset{\,l}) = \Ji(L) \,\cap \downset{\,l}$ and, for
  $\set{j} \cup C \subseteq \Ji(\downset{\,l})$, the relation
  $j \mcovered C$ holds in $\downset{\,l}$ if and only if it holds in
  $L$.
\end{lemma}
\begin{proof}
  Each element of $\Ji(L) \,\cap \downset{\,l}$ is completely \jirr in
  $\downset{\,l}$.  If $x \leq l$, then $x = \bigvee J$ with
  $J \subseteq \Ji(L)$ and clearly $J \subseteq \,\downset{\,l}$.
  Therefore $\downset{\,l}$ is spatial with
  $\Ji(\downset{\,l}) = \Ji(L) \,\cap \downset{\,l}$.

  Suppose now that $\set{j} \cup X \subseteq\, \downset{\,l}$ and
  $j \leq \bigvee X$. If the relations $j \mcovered C$ and
  $C \refines X$ hold in $L$, then $C \subseteq \downset{\,l}$, so
  they hold in $\downset{\,l}$ as well. In particular, this shows that
  $\downset{\,l}$ is \pperfect.
\end{proof}

Let $L$ be a \pperfect lattice. A subset $A \subseteq \Ji(L)$ is
\emph{$D$-closed} if $j \in A$ and $j \mcovered C$ implies $C \subseteq A$.
Given a $D$-closed subset $A \subseteq \Ji(L)$, let $L_{A}$ be the
closure of $A$ under possibly infinite joins so, in particular,
$L_{A}$ is a sub-\jsl of $L$. As $L_{A}$ has infinite joins, it has
also infinite meets. Let us define then $\pi_{A} : L \rto L_{A}$ by
$\pi_{A}(l) := \bigvee \set{ x \in L_{A} \mid x \leq l}$.  The
following Lemma generalizes to \pperfect lattices well known facts
about finite lattices, see e.g. \cite[Lemma 2.33]{FJN}.
\begin{lemma}
  \label{lem:quotients}
  $\pi_{A} : L \rto L_{A}$ is a surjective lattice
  homomorphism. Moreover, $L_{A}$ is a \pperfect lattice whose
  OD-graph is the restriction to $A$ of the OD-graph of $L$.
\end{lemma}
\begin{proof}
  $L_{A}$ is subset of $L$ closed under arbitrary joins and therefore
  $\pi_{A} : L \rto L_{A}$, defined by
  $\pi_{A}(l) := \bigvee \set{ x \in L_{A} \mid x \leq l}$, is a
  surjective map which preserves arbitrary meets (since meets are
  computed in $L_{A}$ via this map, e.g.
  $x \land_{L_{A}} y = \pi_{A}(x \land_{L} y)$).

  Let us show that $\pi_{A}$ preserves arbitrary joins as well. To
  this end, observe first that
  $\pi_{A}(l) = \bigvee \set{ j \in A \mid j \leq l}$.  Since
  $\pi_{A}$ is \op, we only need to show that
  $\pi_{A}(\bigvee X) \leq \bigvee \pi_{A}(X)$. Let therefore
  $j \in A$ with $j \leq \bigvee X$, so $j \mcovered C$ with
  $C \refines X$. Since $j \in A$ and $A$ is $D$-closed, we have
  $C \subseteq A$, whence $C = \pi_{A}(C) \refines \pi_{A}(X)$. It
  follows that $j \leq \bv C \leq \bv \pi_{A}(X)$.

  The set $A \subseteq \Ji(L)$ generates $L_{A}$ under arbitrary joins
  and, moreover, each element of $A$ is completely \jirr in $L_{A}$,
  since $L_{A}$ is a sub-\jsl of $L$; thus $L_{A}$ is spatial and
  $\Ji(L_{A}) = A$. It is easily verified that, for each $j \in A$,
  each \mjc of $j$ $L$ is also a \mjc of $j$ in $L_{A}$.
\end{proof}

  \begin{lemma}
    \label{lem:subdirect}
    If $\set{A_{i} \mid i \in I}$ is a collection of $D$-closed
    subsets such that $\bigcup A_{i} = \Ji(L)$, then
    $\langle \pi_{A_{i} }\mid i \in I \rangle : L \rto \prod
    L_{A_{i}}$ is a lattice embedding, that is, a subdirect
    decomposition of $L$.
  \end{lemma}
  \begin{proof}
    If $l \not\leq l'$, then, by spatiality, there is $j \in \Ji(L)$
    such that $j \leq l$ but $j \not\leq l'$. Let $i \in I$ such that
    $j \in A_{i}$: then $j \leq \pi_{A_{i}}(l)$ but
    $j \not\leq \pi_{A_{i}}(l')$.  It follows that
    $\langle \pi_{A_{i} }\mid i \in I \rangle $ is an injective map.
  \end{proof}

\section{The relational lattices $\R(D,A)$}
\label{sec:rellattices}

Throughout this paper we shall use the notation $\expo{X}{Y}$ for the
set of functions of domain $Y$ and codomain $X$, for $X$ and $Y$ any
two sets.

\medskip

Let $A$ be a collection of attributes (or column names) and let $D$ be
a set of cell values. A \emph{relation} on $A$ and $D$ is a pair
$(\alpha,T)$ where $\alpha \subseteq A$ and
$T \subseteq \expo{\alpha}{D}$.  We shall use $\R(D,A)$ to denote the
set of relations on $A$ and $D$. 
Informally, a relation $(\alpha,T)$ represents a table of a relational
database, with $\alpha$ being the header, i.e. the collection of names
of columns, while $T$ is the collection of rows.

Before we define the \nj, the \iu operations, and the order on
$\R(D,A)$, let us recall some key operations. If
$\alpha \subseteq \beta \subseteq A$ and $f \in \expo{\beta}{D}$, then
we shall use $f \restr[\alpha] \in \expo{\alpha}{D}$ for the
restriction of $f$ to $\alpha$; if $T \subseteq \expo{\beta}{D}$, then
$T \rrestr[\alpha]$ shall denote projection to $\alpha$, that is, the
direct image of $T$ along restriction,
$T \rrestr[\alpha] := \set{ f \restr[\alpha] \mid f \in T}$; if
$T \subseteq \expo{\alpha}{D}$, then $i_{\beta}(T)$ shall denote
cylindrification to $\beta$, that is, the inverse image of
restriction,
$i_{\beta}(T) := \set{ f \in \expo{\beta}{D} \mid f_{\restriction
    \alpha} \in T}$.  Recall that $i_{\beta}$ is \ra to
$\rrestr[\alpha]$.  With this in mind, the \nj $\njoin$ and the inner
union $\iunion$ of relations are respectively described by the
following formulas:
\begin{align*}
  (\alpha_{1},T_{1}) \njoin (\alpha_{2},T_{2})
  & := (\alpha_{1} \cup \alpha_{2},T)  \\
  \text{where }T & = \set{f \mid f \restr[\alpha_{i}] \in T_{i}, i =
    1,2} \\
  & = i_{\alpha_{1} \cup \alpha_{2}}(T_{1}) \cap i_{\alpha_{1} \cup
    \alpha_{2}}(T_{2})\,,  \\
  (\alpha_{1},T_{1}) \iunion (\alpha_{2},T_{2})
  & := (\alpha_{1} \cap \alpha_{2},T) \\
  \text{where }T & = \set{f \mid \exists i\in \set{1,2},\exists
    g \in T_{i} \tst g\, \restr[\alpha_{1} \cap \alpha_{2}] = f} \\
  & = T_{1} \rrestr[\alpha_{1} \cap \alpha_{2}] \cup \,T_{2}
  \rrestr[\alpha_{1} \cap \alpha_{2}]\,.
\end{align*}
The set $\RDA$ is then ordered as follows:
\begin{align*}
  (\alpha_{1},T_{1}) & \leq (\alpha_{2},T_{2}) \quad \tiff \quad \alpha_{2} \subseteq \alpha_{1}
  \tand T_{1} \rrestr[\alpha_{2}] \subseteq T_{2}\,.
\end{align*}
\begin{proposition}[Tropashko \cite{Tropashko2005}]
  The poset $(\RDA,\leq)$ is a lattice, with $\njoin$ as the meet
  operation and $\iunion$ as the join operation. 
\end{proposition}
We shall therefore use $\RDA$ to denote such a lattice and call it a
\emph{relational lattice}. Let us remark however that in
\cite{LitakMHjlamp} such a lattice is called \emph{full} relational
lattice and that the wording ``class of relational lattices'' is used
there for the class of lattices that have an embedding into some
lattice of the form $\R(D,A)$. As our concerns are lattice
theoretical, we shall avoid to use the symbols $\njoin$ and $\iunion$,
and prefer instead the usual meet and join symbols $\land$ and $\vee$.

\medskip

A convenient way of describing these lattices was introduced in
\cite[Lemma 2.1]{LitakMHjlamp}. The authors argued that the relational
lattices $\R(D,A)$ are isomorphic to the lattices of closed subsets of
$A \cup \AD$, where $Z \subseteq A \cup \AD$ is said to be closed if
it is a fixed-point of the closure operator $\closure{(\,-\,)}$
defined as
\begin{align*}
  \closure{Z} & := Z \cup \set{f \in \AD \mid A \setminus Z \subseteq
    Eq(f,g), \text{ for some $g \in Z$} }\,,
\end{align*}
where in the formula above $Eq(f,g)$ is the equalizer of $f$ and
$g$. Letting
\choosedisplay{$\d(f,g) := \set{x \in A \mid f(x) \neq g(x)}$,}{
\begin{align*}
  \d(f,g) & := \set{x \in A \mid f(x) \neq g(x)}\,,
\end{align*}
}
the above definition of the closure operator is obviously equivalent
to the following one:
\begin{align*}
  \closure{Z} & := \alpha \cup \set{f \in \AD \mid \d(f,g) \subseteq
    \alpha, \text{ for some $g \in (Z \cap \AD)$} },\;
  \text{with $\alpha = Z \cap A$}.
\end{align*}
From now on, we rely on this representation of relational lattices.
Relational lattices are atomistic \pperfect lattices. The completely
\jirr elements of $\R(D,A)$ are the singletons $\set{a}$ and
$\set{f}$, for $a \in A$ and $f \in \AD$, see \cite{LitakMHjlamp}. By
an abuse of notation we shall write $x$ for the singleton $\set{x}$,
for $x \in A \cup \AD$. Under this convention, we have therefore
$\Ji(\RDA) = A \cup \AD$.
Every $a \in A$ is \jp, while the \mjc{s} are of the form 
\choosedisplay{$f \mcovered \d(f,g) \cup \set{g}$, }{
  \begin{align*}
    f & \mcovered \d(f,g) \cup \set{g}\,
  \end{align*}
}
for each $f,g \in \AD$, see \cite{San2016}.
The only non-trivial result from \cite{San2016} that we use later (for
Theorem~\ref{thm:embeddings} and Lemma~\ref{lemma:jreducible}) is the
folllowing:
\begin{lemma}
  \label{lemma:noprimecovers}
  Let $L$ be a finite atomistic lattice in the variety generated by
  the class of relational lattices. If
  $\set{j} \cup X \subseteq \Ji(L)$, $j \leq \bigvee X$, and all the
  elements of $X$ are \jp, then $j$ is \jp.
\end{lemma}
The Lemma---which is an immediate consequence of Theorem~7 in
\cite{San2016}---asserts that a join-cover of an element
$j \in \Ji(L)$ which is not \jp cannot be made of \jp elements only.

\section{Overview and statement of the results}
\label{sec:overview}

For an arbitrary frame $\FF$, we  construct in
Section~\ref{sec:onedirection} a lattice $\L(\FF)$; if $\FF$ is
\rooted and full, then $\L(\FF)$ is a subdirectly irreducible lattice,
see Proposition~\ref{prop:LFFsi}.  The key Theorem leading to the
undecidability results is the following one.
\begin{theorem}
  \label{thm:reduction}
  Let $A$ be a finite set and let $\FF$ be an \Sfour \fif
  $A$-frame. There is a surjective $p$-morphism from a universal
  $\Sfive^{A}$-product frame $\U$ to $\FF$ if and only if $\L(\FF)$
  embeds into some relational lattice $\R(D,B)$.
\end{theorem}
\begin{myproof}{outline}
  The construction $\L$ defined in Section~\ref{sec:onedirection}
  extends to a contravariant functor, so if $\U$ is a universal
  $\Sfive^{A}$-product frame and $\psi : \U \rto \FF$ is a surjective
  \pmorphism, then we have an embedding $\L(\psi)$ of $L(\FF)$ into
  $\L(\U)$. We can assume that all the components of $\U$ are equal,
  i.e. that the underlying set of $\U$ is of the form
  $\prod_{a \in A} X$; if this is the case, then $\L(\U)$ is
  isomorphic to the relational lattice $\R(X, A)$.
  
  The converse direction, developed from Section~\ref{sec:ultrametric}
  up to Section~\ref{sec:converse}, is subtler. Considering that
  $\L(\FF)$ is \si, we argue in Section~\ref{sec:intervals} that if
  $\psi : \L(\FF) \rto \R(D,B)$ is a lattice embedding, then we can
  suppose it 
  preserves bounds; in this case $\psi$ has a surjective \la
  $\mu : \R(D,B) \rto \L(\FF)$.
  Let us notice that there is no general reason for $\psi$ to be the
  image by $\L$ of a \pmorphism.  Said otherwise, the functor $\L$ is
  not full and, in particular, the image of an atom by $\mu$ might not
  be an atom. The following considerations, mostly developed in
  Section~\ref{sec:converse}, make it possible to extract a \pmorphism
  from the \la $\mu$.  Since both $\L(\FF)$ and $\R(D,B)$ are
  generated (under possibly infinite joins) by their atoms, each atom
  $x \in \L(\FF)$ has a preimage $y \in \R(D,B)$ which is an atom.
  The set $\Fz$ of \njp atoms of $\R(D,B)$ such that $\mu(f)$ is a
  \njp atom of $\L(\FF)$ is endowed with a $P(A)$-valued distance
  $\d$.  The pair $(\Fz,\d)$ is shown to be a \Pc ultrametric space
  over $P(A)$. Section~\ref{sec:ultrametric} recalls and develops some
  observations on ultrametric spaces valued on powerset algebras. The
  key ones are Theorems~\ref{thm:SecPC} and~\ref{thm:UPC}, stating
  that---when $A$ is finite---\Pc ultrametric spaces over $P(A)$ and
  universal $\Sfive^{A}$-product frames are essentially the same
  objects.  The restriction of $\mu$ to $\Fz$ yields then a surjective
  \pmorphism from $\Fz$, considered as a universal
  $\Sfive^{A}$-product frame, to $\FF$.
\end{myproof}

The following problem was shown to be undecidable in~\cite{HHK2002}:
given a finite $3$-frame $\FF$, does there exists a surjective
$p$-morphism from a universal $\Sfive^{3}$-product frame $\U$ to
$\FF$?  In the introduction we referred to this problem as the
coverability problem of a $3$-frame by a universal
$\Sfive^{3}$-product frame.  The problem was shown to be undecidable
by means of a reduction from the representability problem of finite
simple relation algebras, shown to be undecidable in \cite{HH2001}.
We need to strengthen the undecidability result of \cite{HHK2002} with
some additional observations---rootedness and fullness---as stated in
the following Proposition.
\begin{proposition}
  \label{prop:spmorphundecidable}
  It is undecidable whether, given a finite set $A$ with
  $\card A \geq 3$ and an \Sfour \emph{\fif} $A$-frame $\FF$, there is
  a surjective $p$-morphism from a universal $\Sfive^{A}$-product $\U$
  to $\FF$.
\end{proposition}
\begin{proof}
  Throughout this proof we assume a minimum knowledge of the theory of
  relation algebras, see e.g. \cite{maddux2006}.
  
  The Proposition actually holds if we restrict to the case when
  $\card A = 3$.  Given a finite simple relation algebra $\Afrack$,
  the authors of \cite{HHK2002} construct a $3$-multimodal frame
  $\FF_{\Afrack,3}$ such that $\Afrack$ is representable if and only
  if $\FF_{\Afrack,3}$ is a \pmorphic image of some universal
  $\Sfive^{3}$-product frame. The frame $\FF_{\Afrack,3}$ is \Sfour
  and \rooted \cite[Claim 8]{HHK2002}. We claim that $\FF_{\Afrack,3}$
  is also full, unless $\Afrack$ is the two elements Boolean
  algebra. To prove this claim, let us recall first that an element of
  $\FF_{\Afrack,3}$ is a triple $(t_{0},t_{1},t_{2})$ of atoms of
  $\Afrack$ such that $t_{2}^{\conv} \leq t_{0} ; t_{1}$; moreover, if
  $t,t'$ are two such triples and $i \in \set{0,1,2}$, then
  $t R_{i} t'$ if and only if $t$ and $t'$ coincide in the $i$-th
  coordinate. If $a$ is an atom of $\Afrack$, then $a \leq e_{l} ; a$
  and $a \leq a ; e_{r}$ for two atoms $e_{l},e_{r}$ below the
  multiplicative unit of $\Afrack$.  Therefore, the triples
  $t := (e_{l},a,a^{\conv})$ and $t' = (a,e_{r},a^{\conv})$ are
  elements of $\FF_{\Afrack,3}$ and $tR_{2}t'$. If, for each atom $a$,
  these triples are equal, then every atom of $\Afrack$ is below the
  multiplicative unit, which therefore concides with the top element
  $\top$; since $\Afrack$ is simple, then relation
  $\top = \top ; x ; \top$ holds for each $x \neq \bot$.  It follows
  that $x = \top ; x ; \top = \top$, for each $x \neq \bot$, so
  $\Afrack$ is the two elements Boolean algebra.
    Thus, if $\Afrack$ has more than two elements, then $t \neq t'$
    and $tR_{2}t'$ for some $t,t' \in \FF_{\Afrack,3}$.  Using the
    cycle law of relation algebras, one also gets pairs of distinct
    elements of $\FF_{\Afrack,3}$, call them $u,u'$ and $w,w'$, such
    that $u R_{0} u'$ and $w R_{1} w'$.

    Therefore, if we could decide whether there is a \pmorphism from
    some universal $\Sfive^{3}$-frame to a given \Sfour \fif frame
    $\FF$, then we could also decide whether a finite simple relation
    algebra $\Afrack$ is representable, by answering positively if
    $\Afrack$ has exactly two elements and, otherwise, by answering
    the existence problem of a \pmorphism to $\FF_{\Afrack,3}$.
  \end{proof}

Combining Theorem~\ref{thm:reduction} with
Proposition~\ref{prop:spmorphundecidable}, we derive the following
undecidability result.
\begin{theorem}
  \label{thm:embeddability}
  It is not decidable whether a finite subdirectly irreducible atomistic
  lattice embeds into a relational lattice.
\end{theorem}
Let us remark that Theorem~\ref{thm:embeddability} partly answers
Problem 7.1 in \cite{LitakMHjlamp}.

\smallskip

In \cite{LitakMHjlamp} the authors proved that the quasiequational
theory of relational lattices (i.e. the set of all definite Horn
sentences valid in relational lattices) in the signature
$(\land,\vee,H)$ is undecidable. Here $H$ is the header constant,
which is interpreted in a relational lattice $\R(D,A)$ as the closed
subset $A$ of $A \cup \AD$.  Problem~4.10 in \cite{LitakMHjlamp} asks
whether the quasiequational theory of relational lattices in the
restricted signature $(\land,\vee)$ of pure lattice theory is
undecidable as well. We positively answer this question.
\begin{theorem}
  \label{thm:undecidability}
  The quasiequational theory of relational lattices in the pure
  lattice signature is undecidable.
\end{theorem}

It is a general fact that if the embeddability problem of
finite \si algebras in a class ${\cal K}$ is undecidable, then the
quasiequational theory of ${\cal K}$ is undecidable as well.
We thank one of the authors of \cite{hirsch2012} for pointing out to
us how this can be derived from Evans' work \cite{Evans53} as well as
the connection to their work.  We add here the proof of this fact,
since we shall need it later in the proof of
Theorem~\ref{thm:finiteinfinite}.
\begin{proof}[Proof of Theorem~\ref{thm:undecidability}]
  \newcommand{\ca}{\hat{a}}
  \newcommand{\ba}{\bar{a}}
  Given a
  finite \si algebra $A$ with least non trivial congruence
  $\theta(\ca,\ba)$, we construct a quasiequation $\phi_{A}$ with the
  following property: for any other algebra (in the same signature)
  $K$, $K \not\models \phi_{A}$ if and only if $A$ has an embedding
  into $K$.

  The construction is as follows. Let
  $X_{A} = \set{x_{a} \mid a \in A}$ be a set of variables in
  bijection with the elements of $A$.  For each function symbol $f$ in
  the signature $\Omega$, let $T_{A,f}$ be its table, that is the
  formula
  \begin{align*}
    T_{A,f} & = \bigwedge_{(a_{1},\ldots ,a_{ar(f)}) \in A^{ar(f)}} f(x_{a_{1}},\ldots
    ,x_{ar(f)}) = x_{f(a_{1},\ldots ,a_{ar(f)})}.
  \end{align*}
  We let $\phi_{A}$ be the universal closure of
  $\bigwedge_{f \in \Omega} T_{A,f} \Rightarrow x_{\ca} =
  x_{\ba}$.
  We prove next that an algebra $K$ sastifies $\phi_{A}$ if and only
  if there is no embedding of $A$ into $K$.

  If $K \models \phi_{A}$ and $\psi : A \rto K$, then
  $v(x_{a}) = \psi(a)$ is a valuation such that
  $K,v \models \bigwedge_{f \in \Omega} T_{A,f}$, so
  $\psi(\ca) =v(x_{\ca}) = v(x_{\ba}) = \psi(\ba)$ and $\psi$
  is not injective.

  Conversely, suppose $K \not\models \phi_{A}$ and let $v$ be a
  valuation such that $K,v \models \bigwedge_{f \in \Omega} T_{A,f}$
  and $K,v \not\models x_{\ca} = x_{\ba}$. Define
  $\psi : A \rto K$ as $\psi(a) = v(x_{a})$, then $\psi$ is a
  morphism, since $K,v \models T_{A,f}$ for each $f \in \Omega$.
  Let $Ker_{\psi} = \set{(a,a') \mid \psi(a) = \psi(a')}$ so,
  supposing that $\psi$ is not injective,  $Ker_{\psi}$ is a
  non-trivial congruence. Then
  $(\ca,\ba) \in\theta(\ca,\ba) \subseteq Ker_{\psi}$, so
  $v(x_{\ca}) = \psi(\ca)= \psi(\ba) = v(x_{\ba})$, \contr.
  We have therefore $Ker_{\psi} = \set{(a,a) \mid a \in A}$, which
  shows that $\psi$ is injective.

  Let now ${\cal K}$ be a class of algebras in the same signature. We
  have then
  \begin{align*}
    {\cal K} \not\models \phi_{A} & \tiff K\not\models \phi_{A}
    \text{ for some $K \in {\cal K}$ }\\
    & \tiff \text{there is an embedding  of $A$ into $K$, for some $K \in {\cal K}$ }\,.
  \end{align*}
  Thus, if the embeddability problem of finite \si algebras into some
  algebra in ${\cal K}$ is undecidable, then the quasiequational
  theory of ${\cal K}$ is undecidable as well.
\end{proof}

Following \cite{HHK2002}, let us add some further observations on the
quasiequational theory of relational lattices.
\begin{lemma}
  \label{lemma:ultraproducts}
  The class of lattices that have an embedding into a relational
  lattice is closed under ultraproducts.
\end{lemma}
\begin{proof}
  Let us say that a sublattice $L$ of a lattice $\R(D,A)$ is
  $H$-closed if the subset $A$ belongs to $L$. Let ${\cal R}$ denote
  the closure under isomorphisms of the class of $H$-closed
  sublattices of some $\R(D,A)$.  It is proved in \cite[Corollary
  4.2]{LitakMHjlamp} that ${\cal R}$ is closed under ultraproducts. It
  immediately follows from this result that the class of lattices that
  have an embedding into some relational lattice is closed under
  ultraproducts, as follows.  
  Let $\set{L_{i} \rto \R(D_{i},A_{i})\mid i \in I}$ be a family of
  lattice embeddings and let ${\cal F}$ be an ultrafilter over
  $I$. The ultraproduct constructions on $\set{L_{i} \mid i \in I }$
  and $\set{\R(D_{i},A_{i}) \mid i \in I }$ yield a lattice embedding
  $\ultraprod{L_{i}} \rto \ultraprod{\R(D_{i},A_{i})}$. Clearly, each
  $\R(D_{i},A_{i})$ belongs to ${\cal R}$, whence the ultraproduct
  $\ultraprod{\R(D_{i},A_{i})}$ belongs to ${\cal R}$ as well: thus
  $\ultraprod{\R(D_{i},A_{i})}$ embeds into some $\R(D,A)$, and so
  does $\ultraprod{L_{i}}$.
\end{proof}

\begin{theorem}
  The quasiequational theory of relational lattices is not finitely axiomatizable.
\end{theorem}
\begin{proof}
  A known result in universal algebra---see
  e.g. \cite[Theorem~2.25]{Burris2012}---states that a \si algebra
  satisfies all the quasiequations satisfied by a class of algebras if
  and only if it embeds in an ultraproduct of algebras in this
  class. 
  Lemma~\ref{lemma:ultraproducts} 
  implies that the class of lattices that have an embedding into an
  ultraproduct of relational lattices and the class of lattices that
  have an embedding into some relational lattices are the same. 
  Therefore a \si lattice $L$ embeds in a relational lattice if and
  only if it satisfies all the quasiequations satisfied by the
  relational lattices.  If this collection of quasiequations was a
  logical consequence of a finite set of quasiequations, then we could
  decide whether a finite \si $L$ satisfies all these quasiequations,
  by verifying whether $L$ satisfies the finite set of quasiequations.
  In this way, we could also decide whether such an $L$ embeds into
  some relational lattice.
\end{proof}

Finally, the following Theorem, showing that the quasiequational
theory of the finite relational lattices is stronger than the
quasiequational theory of all the relational lattices, partly answers
Problem~3.6 in \cite{LitakMHjlamp}.
\begin{theorem}
  \label{thm:finiteinfinite}
  There is a quasiequation which holds in all the finite relational
  lattices which, however, fails into an infinite relational lattice.
\end{theorem}
\begin{proof}
  In the first appendix of \cite{HHK2002} an \Sfour \fif $3$-frame
  $\FF$ is constructed that has no surjective \pmorphism from a finite
  universal $\Sfive^{3}$-product frame, but has such a \pmorphism from
  an infinite one.

  Since $\L(\FF)$ is finite whenever $\FF$ is finite, we obtain by
  using Theorem~\ref{thm:reduction} a \si finite lattice $L$ which
  embeds into an infinite relational lattice, but has no embedding
  into a finite one.

  Let $\phi_{L}$ be the quasiequation as in the proof of
  Theorem~\ref{thm:undecidability}. We have therefore that, for any
  lattice $K$, $K \models \phi_{L}$ if and only if $L$ does not
  embed into $K$.
  
  Correspondingly, any finite relational lattice
  satisfies $\phi_{L}$ and, on the other hand,
  $K \not\models \phi_{L}$ if $K$ is the infinite lattice into which
  $L$ embeds.
\end{proof}

\section{The lattice of a multimodal frame}
\label{sec:onedirection}

We assume throughout this Section that $A$ is a finite set of actions.
Given an $A$-frame
$\FF = \langle \X ,\set{R_{a} \mid a \in A} \rangle$, we construct a
lattice 
as follows.  For $\alpha \subseteq A$, we say that $Y \subseteq \X$ is
\emph{$\alpha$-closed} if $x \in Y$, whenever there is a $\alpha$-path
from $x$ to some $y \in Y$.  We say that a subset
$Z \subseteq A \cup \X$ is \emph{closed} if $Z \cap \X$ is $Z \cap A$-closed.
\begin{lemma}
  The collection of closed subsets of $A \cup X$ is a Moore family. 
\end{lemma}
The Lemma, whose proof is straightforward, allows us to define the
lattice of an $A$-frame.

\begin{definition}
  The lattice $\L(\FF)$ is the lattice of closed subsets of
  $A \cup \X$.
\end{definition}

The lattice operations on $\L(\FF)$ are defined as in the
display~\eqref{eq:opsMooreFamily}.
In order to master the formula for the join, we need a more explicit
description of the closure operator associated to this Moore
family. If $\alpha \subseteq A$ and $Y \subseteq \X$, define
\choosedisplay{$\closure[\alpha]{Y} := \set{x \in \X \mid \exists y
    \in Y, x \step[\alpha] y}$.}{
\begin{align*}
  \closure[\alpha]{Y} & := \set{x \in \X \mid \exists y \in Y, x
    \step[\alpha] y}\,.
\end{align*}
}
\begin{lemma}
  For $Z \subseteq A \cup \X$, we have
  \choosedisplay{$\closure{Z} = \alpha \cup \closure[\alpha]{Z \cap
      \X}$, where $\alpha = Z \cap A$.}{
  \begin{align}
    \closure{Z} & = \alpha \cup \closure[\alpha]{Z \cap \X}\,,
    \quad\text{where $\alpha = Z \cap A$.}
    \label{eq:closure}
  \end{align}
  }
\end{lemma}
In particular, for $x \in \X$, $x \in \closure{Z}$ if and only if
there exists $y \in Z \cap \X$ and an $\alpha$-path from $x$ to $y$,
with $\alpha = Z \cap A$.

The above formula~\eqref{eq:closure} allows to make $\L(-)$ into a
contravariant functor from the category of frames to the category of
lattices. Namely, for a $p$-morphism $\psi : \FF_{0} \rto \FF_{1}$ and
any $Z \subseteq A \cup \X[1]$, define
\choosedisplay{$\L(\psi)(Z) := (Z \cap A) \cup \psi^{-1}(Z \cap
  \X[1])$.}{
\begin{align*}
  \L(\psi)(Z) & := (Z \cap A) \cup \psi^{-1}(Z \cap \X[1])\,.
\end{align*}
}
Let $\psi^{A} : A \cup \X[0] \rto A \cup \X[1]$ be the function such
that $\psi^{A}(a) = a$, for each $a \in A$, and
$\psi^{A}(x) = \psi(x)$, for each $x \in \X[0]$. Notice that
$\L(\psi)$ is the inverse image of $\psi^{A}$, so in particular
$\L(\psi)$ commutes with intersections and unions.
\begin{proposition}
  $\L(\psi)$ sends closed subsets of $A \cup \X[1]$ to closed subsets
  of $A \cup \X[0]$. Its restriction to $\L(\FF_{1})$ yields a \bp
  lattice morphism $L(\psi) : \L(\FF_{1}) \rto \L(\FF_{0})$.
\end{proposition}
\begin{proof}
  The key observation is that, for each $\alpha \subseteq A$ and each
  $Y\subseteq \X[1]$, we have $\psi^{-1}(\closure[\alpha]{Y}) =
  \closure[\alpha]{\psi^{-1}(Y)}$:
  \begin{align*}
    \psi^{-1}(\closure[\alpha]{Y}) & = \psi^{-1}(\set{x \in \X[1]\mid
      \exists y \in Y,\, x \step[\alpha] y}) \\
    & = \set{x \in \X[0] \mid \exists y \in Y,\, \psi(x) \step[\alpha] y} \\
    & = \set{x \in \X[0] \mid \exists z \in \psi^{-1}(Y),\, x
      \step[\alpha] z} \tag*{since $\psi$ is a \pmorphism,}
    \\
    &= \closure[\alpha]{\psi^{-1}(Y)}\,.
  \end{align*}    
  This implies that, for $Z \subseteq A \cup \X[1]$, we have
  \choosedisplay{$\L(\psi)(\closure{Z}) = \closure{\L(\psi)(Z)}$.}{
  \begin{align*}
    \L(\psi)(\closure{Z}) & = \closure{\L(\psi)(Z)}\,.
  \end{align*}
  }
 In particular, if $Z \subseteq A \cup \X[1]$ is closed,
  then
  \choosedisplay{$\L(\psi)(Z) = \L(\psi)(\closure{Z}) = \closure{\L(\psi)(Z)}$}{
  \begin{align*}
    \L(\psi)(Z) & = \L(\psi)(\closure{Z}) = \closure{\L(\psi)(Z)}
  \end{align*}
  }
  so $\L(\psi)$ sends closed subsets to closed subsets. $\L(\psi)$
  preserves all meets, since it commutes with intersections. Moreover
  \choosedisplay{
    $\L(\psi)(\bigvee_{i \in I} Z_{i}) = \L(\psi)(\closure{\bigcup_{i
        \in I} Z_{i}}) = \closure{\bigcup_{i \in I} \L(\psi)(Z_{i})} =
    \bigvee_{i \in I} \L(\psi)(Z_{i})$, }{
  \begin{align*}
    \L(\psi)(\bigvee_{i \in I} Z_{i}) & = \L(\psi)(\closure{\bigcup_{i
        \in I} Z_{i}}) = \closure{\bigcup_{i \in I} \L(\psi)(Z_{i})} =
    \bigvee_{i \in I} \L(\psi)(Z_{i})\,,
  \end{align*}
  }
  so $\L(\psi)$ is a lattice morphism.
\end{proof}
As $\L(\psi)$ is the restriction of the inverse image of $\psi^{A}$
defined above, it immediately follows that $\L$ is a contravariant
functor from the category of $A$-frames to the category of lattices. 
\begin{aiml}
  Moreover, we have:
\end{aiml}
\begin{lemma}
  \label{lemma:epictoembedding}
  If $\psi : \FF_{0} \rto \FF_{1}$ is surjective, then $\L(\psi)$ is
  injective.
\end{lemma}
\begin{proof}
  If $\psi$ is surjective, then $\psi^{A}$ is also surjective. As
  $\L(\psi)$ is the inverse image of $\psi^{A}$, then $\L(\psi)$ is
  injective.
\end{proof}

We are ready to state the main result of this Section.
\begin{theorem}
  \label{thm:direct}
  If there exists a surjective \pmorphism from a universal
  $\Sfive^{A}$-product frame $\U$ to an $A$-frame $\FF$, then
  $\L(\FF)$ embeds into a relational lattice.
\end{theorem}
\begin{proof}
  We say that $\U$ is uniform on $X$ if all the components of $\U$ are
  equal to $X$. Spelled out, this means that
  $X_{\U} = \prod_{a \in A} X$.
  Let $\psi : \U \rto \FF$ be a \pmorphism as in the statement of the
  Theorem. W.l.o.g. we can assume that $\U$ is uniform on some set
  $X$. If this is not the case, then we choose $a_{0} \in A$ such that
  $X_{a_{0}}$ has maximum cardinality and surjective mappings
  $p_{a} : X_{a_{0}} \rto X_{a}$, for each $a \in A$.  The product
  frame $\U'$ on $\prod_{a \in A} X_{a_{0}}$ is uniform and
  $\prod_{a \in A} p_{a} : \U' \rto \U$ is a surjective \pmorphism. By
  pre-composing $\psi$ with this \pmorphism, we obtain a surjective
  \pmorphism from the uniform $\U'$ to $\FF$.
  Now, if $\U$ is uniform on $X$, then $\L(\U)$ is equal to the
  relational lattice $\R(X,A)$. Then, by functoriality
  of $\L$, we have a lattice morphism
  \choosedisplay{$L(\psi) : \L(\FF) \rto \L(\U) = \R(X,A)$.}{
  \begin{align*}
    \L(\psi) : \L(\FF) \rto \L(\U) = \R(X,A)\,.
  \end{align*}
  }
  By Lemma~\ref{lemma:epictoembedding} $\L(\psi)$ is an embedding.
\end{proof}

\subsection*{Some properties of the lattices $\L(\FF)$}
\begin{proposition}
  The \cjirr elements of $\L(\FF)$ are the singletons, so $\L(\FF)$ is
  an atomistic lattice. 
\end{proposition}
\begin{proof}
  Each singleton set is closed. It immediately follows that the \jirr
  elements of $\L(\FF)$ are the singletons and  clearly they are
  atoms.
\end{proof}

Identifying the singletons of $P(A \cup \X)$ with their elements, we
can write 
\begin{align*}
  \Ji(\L(\FF)) & = A \cup \X.
\end{align*}
To state the next Proposition,
let us say that an $\alpha$-path from $x$ to $y$ is \emph{minimal} if
there is no $\beta$-path from $x$ to $y$, for each proper subset
$\beta$ of $\alpha$.
\begin{proposition}
  \label{prop:minjoincovers}
  $\L(\FF)$ is a \pperfect lattice.  Each element of $A$ is \jp, while
  the \mjc{s} of $x \in \X$ are of the form
  $x \mcovered \alpha \cup \set{y}$, for a minimal $\alpha$-path from $x$ to
  $y$.
\end{proposition}
\begin{proof}
  We observe that, for $Z \subseteq A \cup \X$, the relation
  \begin{align*}
    x & \in \bigvee Z = \closure{Z}
  \end{align*}
  holds if and only if either (i) $x \in A \cap Z$, or (ii) $x \in \X$
  and $x \step[\alpha] y$ for some $y \in \X \cap Z$ and some
  $\alpha \subseteq A \cap Z$. Thus, in particular, each element of
  $A$ is \jp. If $x \in \X$, $Z \subseteq A \cup \X$ and $x \leq \bv Z$,
  then we can find $y \in \X \cap Z$ and an $\alpha$-path from $x$ to
  $y$ with $\alpha \subseteq A \cap Z$.  Clearly, we can assume
  $x \step[\alpha] y$ is minimal, so
  \begin{align*}
    x \in \closure{\alpha \cup \set{y}} = \bigvee \alpha \vee y\,,
  \end{align*}
  with $\alpha \cup \set{y} \subseteq Z$. This proves that every cover
  of $x$ refines to a cover of the form $x \leq \bv \alpha \vee y$
  with $x \step[\alpha] y$ minimal.
\end{proof}
Notice that if an $\alpha$-path is minimal, then $\alpha$ is
necessarily finite. Therefore $\L(\FF)$ is actually a lattice with the
$\Sigma$-weak \mjc refinement property as defined in
\cite{wehrung:MCRP}, where $\Sigma$ is here the set of \cjirr elements
of the lattice.

Before stating the next Proposition, let us recall from
\cite[Corollary 2.37]{FJN}, see also
\cite[Section~5.2]{San09:duality}, that a finite lattice $L$ is \si if
and only if the directed graph $(\Ji(L),D)$ is \rooted.  Here $D$ is
the \emph{join-dependency relation} on the \jirr elements of $L$,
\begin{noaiml}
defined as follows:
\begin{align*}
  j D k & \text{ iff } j \neq k \text{ and, for some $p \in L$, }  
  j \leq p \vee k \text{ and } j \neq p \vee k_{\ast}\,,
\end{align*}
where $k_{\ast}$ denotes the unique lower cover of $k \in \Ji(L)$.  
  It can be shown that $j D k$ if and only if $k \neq j$ and $k \in C$
  for some subset $C \subseteq \Ji(L)$ such that $j \mcovered C$, see
  e.g. \cite[Lemma 2.31]{FJN}.
If a lattice is atomistic, then $k_{\ast} = \bot$ for each
$k \in \Ji(L)$, and therefore $j D k$ if and only if $j \neq k$ and
$j \leq p \vee k$ for some $p \in L$ with $j \not \leq p$.
\end{noaiml}
\begin{aiml}
  which, on atomistic finite lattices, can be defined by saying that
  $j D k$ holds if $j \neq k$ and $j \leq p \vee k$ for some $p \in L$
  with $j \not \leq p$.
\end{aiml}
\begin{proposition}
  \label{prop:LFFsi}
  If a finite $A$-frame $\FF$ is \rooted and full, then $\L(\FF)$ is a
  \si lattice.
\end{proposition}
\begin{proof}
  We argue that the digraph $(\Ji(\L(\FF)),D)$ is \rooted.  Observe
  that $x \in \closure{\set{a,y}} = a \vee y$ whenever $x R_{a}
  y$.
  This implies that $x Dy $ and $xDa$ when $x, y \in \X$, $a \in A$,
  $x \neq y$ and $x R_{a} y$. The fact that of $(\Ji(\L(\FF)),D)$ is
  \rooted follows now from $\FF$ being \rooted and full.
\end{proof}

\section{Some theory of generalized ultrametric spaces}
\label{sec:ultrametric}

Generalized ultrametric spaces over a Boolean algebra $P(A)$ turn out
to be a useful tool for relational lattices
\cite{LitakMHjlamp,San2016}---as well as, we claim here, for universal
product frames from multidimensional modal logic \cite{Kurucz2007}. 
The use of metrics is well known in graph theory, where universal
product frames are known as Hamming graphs, see e.g.
\cite{HIK2012}. 
Generalized ultrametric spaces over a Boolean algebra $P(A)$ were
introduced in \cite{PriessCrampeRibenboim1995} to study equivalence
relations.
The main results of this Section are Theorem~\ref{thm:SecPC} and
Proposition~\ref{prop:USec} which together substantiate the claim that
when $A$ is finite, universal $\Sfive^{A}$-product frames are \Pc
ultrametric spaces valued in the Boolean algebra $P(A)$.  It is this
abstract point of view that shall allow us to construct a universal
product frame given a lattice embedding $\L(\FF) \rto \RDA$.

Some of the observations we shall develop are not strictly necessary
to prove the undecidability result, which is the main result of this
paper; namely, we can always suppose that the set $A$ is finite.
Nonetheless we include these observations since they are part of a
coherent set of results and, as far as we are aware of, they are
original.

\begin{definition}
  An \emph{ultrametric space over $P(A)$} (briefly, a \emph{space}) is a pair
  $(X,\d)$, with $\d : X \times X \rto P(A)$ such that, for every
  $f,g,h \in X$,
  \begin{align*}
    \delta(f,f) & \subseteq \emptyset\,, &
    \delta(f,g) & \subseteq \d(f,h) \cup \d(h,g)\,.
  \end{align*}
\end{definition}
That is, we have defined an ultrametric space over $P(A)$ as a
category (with a small set of objects) enriched over
$(P(A)^{op},\emptyset,\cup)$, see \cite{lawvere}. We shall assume in
this paper that such a space $(X,\d)$ is also \emph{reduced} and
\emph{symmetric}, that is, that the following two properties hold for
every $f,g \in X$:
\begin{align*}
  \d(f,g) & = \emptyset \text{ implies } f = g, & 
  \d(f,g) & =
  \d(g,f)\,.
\end{align*}
Under these hypothesis, it is easily seen that if $A$ is empty or a
singleton, then the categories of spaces over $P(A)$ are
trivial. Thus, we shall assume here that $A$ has at least two
elements.

A \emph{morphism} of spaces\footnote{As $P(A)$ is not totally ordered,
  we avoid calling a morphism ``\emph{non expanding map}'' as it is
  often done in the literature.}  $\psi : (X,\d_{X}) \rto (Y,\d_{Y})$
is a function $\psi: X \rto Y$ such that
$\d_{Y}(\psi(f),\psi(g)) \leq \d_{X}(f,g)$, for each $f,g \in X$. If
$\d_{Y}(\psi(f),\psi(g)) = \d_{X}(f,g)$, for each $f,g \in X$, then
$\psi$ is said to be an \emph{isometry}.
For $(X,\d)$ a space over $P(A)$, $f \in X$ and $\alpha \subseteq A$,
the ball centered in $f$ of radius $\alpha$ is defined as usual:
$B(f,\alpha) := \set{g \in X \mid \delta(f,g) \subseteq \alpha}$. In
\cite{Ackerman2013} a space $(X,\d)$ is said to be \emph{\Pc} if, for
each $f,g \in X$ and $\alpha,\beta \subseteq A$,
\choosedisplay{$B(f,\alpha \cup \beta) = B(g,\alpha \cup \beta)$
  implies $B(f,\alpha) \cap B(g,\beta) \neq\emptyset$. }{
\begin{align*}
  B(f,\alpha \cup \beta) = B(g,\alpha \cup \beta)
  & \text{ implies } B(f,\alpha) \cap B(g,\beta) \neq\emptyset\,.
\end{align*}
}
This  property is easily seen to be equivalent to:
\begin{align*}
  \delta(f,g) \subseteq \alpha \cup \beta& \text{ implies }
  \delta(f,h) \subseteq \alpha \text{ and } \delta(h,g) \subseteq
  \beta\,,
  \;\text{ for some $h \in X$}.
\end{align*}
Recall also from \cite{Ackerman2013} that a 
space is said to be \emph{spherically complete} if the intersection
$\bigcap_{i \in I} B(f_{i},\alpha_{i})$ of every chain
$\set{B(f_{i},\alpha_{i}) \mid i \in I}$ of balls is non-empty. It was
shown in \cite{PriessCrampeRibenboim1995} that, when $A$ is finite,
every space over $P(A)$ is \SC.
To stress the importance of these two conditions, \PC and \SCC, let us
recall the following result from \cite{Ackerman2013}.
\begin{proposition}
  In the category of spaces, the injective objects are the spaces that
  are both \Pc and \SC.
\end{proposition}
If $(X,\d_{X})$ is a space and 
$Y \subseteq X$, then the restriction of $\d_{X}$ to $Y$ induces a
space $(Y,\d_{X})$; we say then that $(Y,\d_{X})$ is a \emph{subspace}
of $X$.  Notice that the inclusion of $Y$ into $X$ yields an isometry
of spaces.

Our main example of space over $P(A)$ is $(\AD,\d)$, with $\AD$ the
set of functions from $A$ to $D$ and the distance defined by
\choosedisplay{$\d(f,g) := \set{ a \in A \mid f(a) \neq g(a)}$.}{
\begin{align}
  \label{def:distance}
  \d(f,g) & := \set{ a \in A \mid f(a) \neq g(a)}\,.
\end{align}
} 
A second example is a slight generalization of the previous one.
Given a surjective function $\pi : E \rto A$, let $\Sec{\pi}$ denote
the set of all sections of $\pi$, that is the functions $f : A \rto E$
such that $\pi \circ f = id_{A}$; the formula in \eqref{def:distance}
also defines a distance on $\Sec{\pi}$.
Clearly, $(\Sec{\pi},\d)$ is a subspace of $(E^{A},\d)$ and,
considering the first projection $\pi_{1} : A \times D \rto A$, we can
see that $(D^{A},\d)$ is isomorphic to the space $(\Sec{\pi_{1}},\d)$.
By identifying $f \in \Sec{\pi}$ with the vector
$\langle f_{a} \in \pi^{-1}(a)\mid a \in A\rangle$, we see that
\begin{align}
  \label{eq:sepuniversalproductframe}
  \Sec{\pi} & = \prod_{a \in A}E_{a}\,, \quad\text{where
    $E_{a} := \pi^{-1}(a)$. }
\end{align}
That is, the underlying set of a space $(\Sec{\pi},\d)$ is that of a
universal $\Sfive^{A}$-product frame. Our next observations are meant
to identify the particular role of the universal $\Sfive^{A}$-product
frames as spaces.

\begin{proposition}
  Every space of the form $(\Sec{\pi},\d)$ is both \Pc and \SC.
\end{proposition}
\begin{proof}
  It is immediate to verify that $(\Sec{\pi},\d)$ is  \Pc.
  For \SCC, let ${\cal C} := \set{B(f_{i},\alpha_{i}) \mid i \in I}$
  be a chain of balls. For each $a \in A$ pick $\ast_{a} \in D_{a}$
  and define $f$ as follows:
  \begin{align*}
    f(a) & =
    \begin{cases}
      f_{i}(a)\,, & \text{if $a \not\in \alpha_{i}$ for some $i \in I$,}
      \\
      \ast_{a} \,, & \text{otherwise.}
    \end{cases}
  \end{align*}
  Let us show that $f$ is well defined. Namely, suppose that
  $a \not\in \alpha_{i}$ and $a \not\in \alpha_{j}$. Since ${\cal C}$
  is a chain, we can suppose, without loss of generality, that
  $B(f_{i},\alpha_{i}) \subseteq B(f_{j},\alpha_{j})$ so
  $\d(f_{i},f_{j}) \subseteq \alpha_{j}$.  Since
  $a \not\in \alpha_{j}$ it follows that $f_{i}(a) = f_{j}(a)$.

  Let now $i \in I$ be arbitrary; if $a \not \in \alpha_{i}$, then
  $f(a) = f_{i}(a)$, so $\d(f,f_{i}) \subseteq \alpha_{i}$ and
  $f \in B(f_{i},\alpha_{i})$. 
  It follows that 
  $f \in \bigcap_{i \in I} B(f,\alpha_{i})$.
\end{proof}

\color{black}

\begin{theorem}
  \label{thm:SecPC}
  Every space $(X,\d)$ over $P(A)$ has an isometry into some space of
  the form $(\Sec{\pi},\d)$. If $(X,\d)$ is pairwise and \SC, then
  this isometry is an isomorphism.
\end{theorem}
\begin{proof}
  For each $a \in A$, let
  $D_{a} = \set{B(f,A \setminus \set{a}) \mid f \in X}$. That is, $D_{a}$ is
  the quotient of $X$ by the equivalence relation identifying $f$ and
  $g$ when $\d(f,g) \subseteq A \setminus \set{a}$. Let
  $\pi : \sum_{a \in A} D_{a} \rto A$ be the obvious projection.
  
  We associate to $f \in X$ the vector $\psi(f) = \langle B(f,A \setminus \set{a})
  \mid a \in A\rangle$.
  Let us argue that the correpondence $\psi$ is an isometry:
  \begin{align*}
    a \not\in \d(\psi(f),\psi(g))
    & \tiff B(f,A \setminus \set{a}) =
    B(g, A \setminus \set{a}) \\
    & \tiff \d(f,g) \subseteq A \setminus \set{a} 
    \tiff  a \not\in \d(f,g)\,,
  \end{align*}
  thus $ \d(\psi(f),\psi(g)) = \d(f,g)$. In particular, when the space
  is reduced (i.e. $\delta(f,g) = \emptyset$ implies $f = g$), $\psi$
  is an injective map.

  \medskip

  \newcommand{\Aplus}[1]{A_{#1^{+}}}
  \newcommand{\Amoins}[1]{A_{#1^{-}}}

  Next, we suppose that $(X,\d)$ is pairwise and \SC and argue that
  $\psi$ is surjective.  To this goal, we fix a well-ordering on $A$,
  say $A = \set{a_{\lambda} \mid \lambda < \tau}$. For
  $\lambda < \tau$, let us also set
  $\Amoins{\lambda} := \set{a_{\beta} \in A \mid \beta \leq \lambda}$
  and
  $\Aplus{\lambda} := A \setminus \Amoins{\lambda} = \set{a_{\beta}
    \in A \mid \beta > \lambda}$.
  
  Let
  $v := \langle B(f_{\lambda},A \setminus \set{a_{\lambda}}) \mid
  \lambda < \tau \rangle \in \Sec{\pi}$;
  we need to construct a preimage of $v$ by $\psi$.
  To this end, we construct, by induction on $\lambda < \tau$, a
  family $\set{g_{\lambda} \in X \mid \lambda < \tau }$ such that
  $g_{\lambda} \in B(g_{\beta},\Aplus{\beta})$ for
  $\beta \leq \lambda$ and
  $\d(g_{\lambda},f_{\lambda}) \subseteq A \setminus
  \set{a_{\lambda}}$.
  Let $\lambda < \tau$ be an ordinal and suppose that we have defined
  $g_{\beta}$ with these properties for each $\beta < \lambda$. As
  $\set{ B(g_{\beta},\Aplus{\beta}) \mid \beta < \lambda}$ is a chain,
  we can pick
  $g \in \bigcap_{\beta < \lambda} B(g_{\beta},\Aplus{\beta})$.
  Notice that if $\lambda = \gamma + 1$ is a successor cardinal, then
  we can simply pick $g_{\gamma}$.

  We use \PC to define $g_{\lambda}$ as some $h$ with
  $\d(g, h) \subseteq \set{a_{\lambda}}$ and
  $\d(h,f_{\lambda}) \subseteq \d(g,f_{\lambda}) \setminus
  \set{a_{\lambda}}$
  (if $a_{\lambda} \not\in \d(g,f_{\lambda})$, then we can take
  $h = g$).
  Clearly,
  $\d(g_{\lambda},g_{\lambda}) = \emptyset \subseteq \Aplus{\lambda}$
  and, for $\beta < \lambda$, we have
  $\delta(g_{\lambda},g_{\beta}) \subseteq
  \delta(g_{\lambda},g) \cup \delta(g,g_{\beta}) =
  \set{a_{\lambda}} \cup \delta(g,g_{\beta}) \subseteq
  \set{a_{\lambda}} \cup \Aplus{\beta} \subseteq \Aplus{\beta}$.

  Let now $g \in \bigcap_{\lambda < \tau} B(g_{\beta},\Aplus{\beta})$.
  If $\lambda < \tau$, then
  \begin{align*}
    \d(g,f_{\lambda})
    & \subseteq \d(g,g_{\lambda}) \cup \d(g_{\lambda},f_{\lambda})
    \subseteq \Aplus{\lambda} \cup (A \setminus \set{a_{\lambda}})
    \subseteq A \setminus \set{a_{\lambda}}\,.
  \end{align*}
  This shows that
  $B(g, A \setminus \set{a_{\lambda}}) = B(f_{\lambda},A \setminus
  \set{a_{\lambda}})$
  or, stated otherwise, $\psi(g)_{\lambda} = v_{\lambda}$, for each
  $\lambda < \tau$, so $g$ is a preimage of $v$.  \color{black}
\end{proof}

By the last two Propositions and the last Theorem, we obtain:
\begin{corollary}
  Universal product frames are, up to isomorphism, the injective
  objects in the category of spaces.
\end{corollary}

\subsection*{Continuous supbspaces and completeness}
In the following, let $(X,\d_{X})$ be a fixed \Pc space. 
Our next goal is to devise criteria to recognize \Pc subspaces of
$(X,\d_{X})$. To this end, let us introduce the notion of continuous
subspaces of $(X,\d_{X})$ as follows: for a subspace $Y$ of $X$, we
define
\begin{align*}
  \vv_{Y}(f) & := \bigcap \,\set{\d(f,g) \mid g \in Y}\,, 
\end{align*}
and say that a subspace $Y$ of $X$ is \emph{\continuous} if, for
each $f \in X$, $\vv_{Y}(f) = \emptyset$ implies $f \in Y$.
\begin{lemma}
  \label{lemma:contimplpc}
  A continuous subspace $Y$ of $X$ is \Pc.
\end{lemma}
\begin{proof}
  Let $f,g \in Y$ with $\d(f,g) \subseteq \alpha \cup \beta$.  Let
  $h \in X$ be such that $\d(f,h) \subseteq \alpha \setminus \beta$
  and $\d(h,g) \subseteq \beta$; then
  $\bigcap_{k \in Y} \d(h,k) \subseteq \d(h,f) \cap \d(h,g) \subseteq
  \alpha \setminus \beta \cap \beta = \emptyset$,
  so $h \in Y$ since $Y$ is \continuous.
\end{proof}
\begin{lemma}
  \label{lemma:contimplpc}
  If $X$ is also \SC, then a continuous subspace $Y$ of $X$ is \PSc.
\end{lemma}
\begin{proof}
  By Theorem~\ref{thm:SecPC}, we can suppose that $X$ is of the form
  $\prod_{a \in A} E_{a}$. A subset $Y$ of $X$ is then continuous if
  $f \in Y$, whenever for all $a \in A$ there exists $g \in Y$ with
  $f(a) = g(a)$. Let us put $Y_{a} := \set{ g(a) \mid g \in Y}$, so
  $Y \subseteq \prod_{a \in A} Y_{a}$. If
  $f \in \prod_{a \in A} Y_{a}$, then for each $a \in A$ there is some
  $g \in G$ such that $f(a) =g(a)$ and therefore $f \in Y$. We have
  therefore $Y = \prod_{a \in A} X_{a}$, so $Y$ is \PSc.
\end{proof}

\begin{lemma}
  \label{lemma:pairwise}
  If $Y$ is \Pc subspace of $X$, then, for each $f \in X$, the set
  $\set{\alpha \subseteq A \mid B(f,\alpha) \cap Y \neq \emptyset}$ is
  closed under finite intersections.  Consequently, if $Y$ is \SC,
  then, for each $f \in X$, there exists $g \in Y$ with
  $\d(f,g) = \vv_{Y}(f)$.
\end{lemma}
\begin{proof}
  Let ${\cal A}$ be the set of all $\alpha \subseteq A$ such that
  $B(f,\alpha) \cap Y \neq \emptyset$.  For each
  $\alpha \in {\cal A}$, choose $t_{\alpha} \in B(f,\alpha) \cap Y$.

  Observe that, for $\alpha, \alpha' \in {\cal A}$,
  $\d(t_{\alpha},t_{\alpha'}) \subseteq \d(t_{\alpha},f) \cup
  \d(f,t_{\alpha'}) \subseteq \alpha \cup \alpha'$.
  That is, the function $t$ sending $\alpha$ to $t_{\alpha}$ is a
  $\gamma$-Cauchy function to $Y$ as defined in \cite[Definition
  2.8]{Ackerman2013}, where $\gamma = \bigcap {\cal A}$.

  Observe next that
  $\d(t_{\alpha},t_{\alpha'}) \subseteq \alpha \cup \alpha'$ so, by
  \PC of $Y$, $\d(t_{\alpha},h) \subseteq \alpha$ and
  $\d(h,t_{\alpha'}) \subseteq \alpha'$ for some $h \in Y$. It follows
  that $\d(f,h) \subseteq \alpha \cap \alpha'$, showing that
  $\alpha \cap \alpha' \in {\cal A}$. In particular, $t$ is a
  $\gamma$-Cauchy net, as defined in \cite[Definition
  2.9]{Ackerman2013}.

  If we assume that $Y$ is \SC, then we can use Proposition 2.16
  in~\cite{Ackerman2013} to deduce that, for some $g \in Y$,
  $\d(g,t_{\alpha}) \subseteq \alpha$, for each $\alpha \in {\cal A}$.
  For such a $g \in Y$, we have
  $\d(f,g) \subseteq \d(f,t_{\alpha}) \cup \d(t_{\alpha},g) \subseteq
  \alpha$,
  showing that $\d(f,g) \subseteq \vv_{Y}(f)$. As $g \in Y$, we also
  have $\vv_{Y}(f)\subseteq \d(f,g)$ and $\vv_{Y}(f) = \d(f,g)$.
\end{proof}

\begin{corollary}
  \label{cor:pcimplcont}
  If $Y$ is a \Pc and \SC subspace of $X$, then $Y$ is a \continuous
  subspace of $X$.  
\end{corollary}
\begin{proof}
  Let $f \in X$ be such that
  $\vv_{Y}(f) = \bigcap_{g \in Y} \d(f,g) = \emptyset$. By
  Lemma~\ref{lemma:pairwise}, let $h \in Y$ such that
  $\vv_{Y}(f) = \d(f,h)$, so $\d(f,h) = \emptyset$ and $f = h \in Y$.
\end{proof}
\begin{corollary}
  If $X$ is also \SC, then a subspace of $X$ is continuous if and only
  if it is \PSc.
\end{corollary}
Let us remark that when $A$ is finite, then $(X,\d_{X})$ and all its
subspaces are \SC, so in this case continuity and \PC are equivalent
conditions.

\medskip

\subsection*{Modules} We say that a function $v : X \rto P(A)$ is a
\emph{module} if
\begin{align*}
  v(f) & \subseteq \d(f,g) \cup v(g)\,.  
\end{align*}
In (enriched) category theory ``module'' is a standard naming for an
enriched functor from an enriched category to the base category
enriched on itself. Here a module can be seen as a space morphism from
$(X,\d)$ to the space $(P(A),\Delta)$, where $\Delta$ is the symmetric
difference.

We let $Modules(X)$ be the set of all modules $v$; we order this set
by letting $v \leq w$ if and only if $w(f) \subseteq v(f)$, for each
$f \in Y$ (that is, we take the reverse pointwise order). Let use
$Sub(X)$ for the set of subspaces of $X$, ordered by inclusion---thus
$Sub(X)$ is the usual power set of $X$.
Given a module $v$, let us define
\choosedisplay{$\S_{v} := \set{x \in X \mid v(x) = \emptyset}$.}{
\begin{align*}
  \S_{v} & := \set{x \in X \mid v(x) = \emptyset}\,.
\end{align*}
}
It is easily seen that $\S : Modules(X) \rto Sub(X)$, sending $v$ to
$\S_{v}$, is also \op.
\begin{lemma}
  The map $\vv$ is \la to $\S$.
\end{lemma}
\begin{proof}
  As both maps are \op, we shall show that the usual unit and
  counit laws hold.
  If $f \in Y$, then $\vv_{Y}(f) \subseteq \d(f,f) = \emptyset$;
  thus $Y \subseteq \S_{\vv_{Y}}$.
  Let us argue for the counit law. For each $f \in X$ and
  $g \in \S_{v}$---i.e. when $v(g) = \emptyset$---we have
  $v(f) \subseteq \d(f,g) \cup v(g) = \d(f,g)$.  It follows that
  $v(f) \subseteq \vv_{\S_{v}}(f)$, for each $f \in X$.  This means
  that $\vv_{\S_{v}} \leq v$ in $Modules(X)$.
\end{proof}
  
\begin{lemma}
  \label{lemma:pairwise2}
  \label{lemma:kernmodulepairwisecomplete}
  For each module $v$, $\S_{v}$ is a \continuous subspace of $X$.
\end{lemma}
\begin{proof}
  We already observed that $v(f) \subseteq \vv_{\S_{v}}(f)$, that is,
  $v(f) \subseteq \bigcap_{g \in \S_{v}} \d(f,g)$. If the latter
  expression is equal to the emptyset, then $v(f) = \emptyset$, whence
  $f \in \S_{v}$. This shows that $\S_{v}$ is a \continuous subspace
  of $X$.
\end{proof}

\begin{proposition}
  \label{prop:pairwiseclosure}
  A subspace $Y$ of $X$ is \continuous if and only
  if $\S_{\vv_{Y}} = Y$.  Thus, for any $Y \subseteq X$,
  $\S_{\vv_{Y}} \subseteq X$ is the least \continuous subspace of $X$
  containing $Y$.
\end{proposition}
\begin{proof}
  Observe first that if $Y =\S_{\vv_{Y}}$, then $Y$ is \continuous by
  Lemma~\ref{lemma:pairwise2}.
  
  Conversely, let us suppose that $Y$ is \continuous. By adjointness,
  $Y \subseteq \S_{\vv_{Y}}$ holds, so we argue for the reverse
  inclusion.  If $f \in \S_{\vv_{Y}}$, then
  $\emptyset = \vv_{Y}(f) = \bigcap_{g \in Y} \d(f,g)$, so $f \in
  Y$. Therefore $\S_{\vv_{Y}} \subseteq Y$.

  The last statement follows from the characterization of the \continuous
  subspaces of $X$ as the fixed-points of the closure operator
  $\S_{\vv(-)}$.
\end{proof}
\begin{proposition}
  A module $v$ is such that $\vv_{\S_{v}} = v$ if and only if either
  $v(f) = A$, for each $f \in X$, or $v(f) =\emptyset$, for some
  $f \in X$.
\end{proposition}
\begin{proof}
  If $v(f) = A$ for each $f \in X$, then $S_{v} = \emptyset$ (since we
  are assuming that $A \neq \emptyset)$. It follows that
  $\vv_{\S_{v}}(f) = A$ for each $f \in X$ and $\vv_{\S_{v}} = v$.
  
  Suppose now that $v(g) = \emptyset$ for some $g \in X$.  By
  adjointness, we have $v(f) \subseteq \vv_{\S_{v}}(f)$, for all
  $f \in X$; thus we need to argue for the opposite inclusion.  Fix
  $f \in X$; we exhibit next $h \in X$ such that $v(h) = \emptyset$
  and $\d(f,h) \subseteq v(f)$. It shall follow that
  $\vv_{\S_{v}}(f) = \bigcap_{v(h) = \emptyset} \d(f,h) \subseteq
  v(f)$.
  Since $v(f) \subseteq \d(f,g) \cup v(g) = \d(f,g)$, we can write
  $\d(f,g) = v(f) \cup (\delta(f,g) \setminus v(f))$. We use now \PC
  to pick $h \in X$ such $\d(f,h) \subseteq v(f)$ and
  $\d(h,g) \subseteq \d(f,g) \setminus v(f)$.  Then
  $v(h) \subseteq \d(h,f) \cup v(f) \subseteq v(f)$, and
  $v(h) \subseteq \delta(h,g) \cup v(g) = \delta(h,g) \subseteq
  \delta(f,g) \setminus v(f)$.
  It follows that $v(h) \subseteq v(f) \cap A \setminus v(f)$, whence
  $v(h) = \emptyset$.

  For the converse direction, suppose that $v(f) = \vv_{\S_{v}}(f)$
  for each $f \in X$. If $v(f) \neq \emptyset$, for each $f \in Y$,
  then $v(f) = \vv_{\S_{v}}(f) = A$, for each $f \in X$.  Otherwise,
  $v(f) = \emptyset$, for some $f \in X$.
\end{proof}

\begin{remark}
  Proposition~\ref{prop:pairwiseclosure}, characterizing \continuous
  subspaces as closed subsets of a closure operator, suggests that \Pc
  spaces might have some algebraic nature as well.
  This is actually the case.  It is easily verified that a space
  $(X,\d)$ is \Pc if and only if, \emph{for each $\alpha,\beta$ such
    that $\alpha \cap \beta =\emptyset$}, and for each $f,g \in X$,
  $\d(f,g) \subseteq \alpha \cup \beta$ implies
  $\d(f,h) \subseteq \alpha$ and $\d(h,g) \subseteq \beta$ for some
  $h \in X$. We observe that \emph{such an $h$ 
    is unique}.
  Suppose that $\alpha \cap \beta = \emptyset$ and let $h_{i}$,
  $i =1,2$ with $\d(f,h_{i}) \subseteq \alpha$ and
  $\d(h_{i},g) \subseteq \beta$.  Then
  $\delta(h_{1},h_{2}) \subseteq \delta(h_{1},f) \cup \delta(f,h_{2})
  \subseteq \alpha$
  and similarly $\delta(h_{1},h_{2}) \subseteq \beta$. It follows that
  $\d(h_{1},h_{2}) \subseteq \alpha \cap \beta = \emptyset$ and
  $h_{1} = h_{2}$.
\end{remark}

\subsection*{Pairwise complete spaces and universal product frames}

We already observed, in the
display~\eqref{eq:sepuniversalproductframe}, that the underlying set
of a space of the form $(\Sec{\pi},\d)$ with $\pi : E \rto A$ is that
of a universal $\Sfive^{A}$-product frame.
Something more is true: we can define the transition relations of the
universal $\Sfive^{A}$-product frame by means of the metric. Indeed,
for each $a \in A$, we have
\choosedisplay{$f R_{a} g \text{ iff } \d(f,g) \subseteq \set{a}$.}{
\begin{align*}
  f R_{a} g & \text{ iff } \d(f,g) \subseteq \set{a}\,.
\end{align*}
} 
On the other hand, if $A$ is finite, then the metric is completely
determined from the transition relation of the frame, using the notion
of $\alpha$-path introduced in Section~\ref{sec:defselconcepts}, as
follows:
\choosedisplay{$\d(f,g) =\bigcap \set{\alpha \subseteq A \mid f
    \xrightarrow{\alpha} g}$.}{
\begin{align*}
  \d(f,g) & =\bigcap \;\set{\alpha \subseteq A \mid \text{there
      exists an $\alpha$-path from $f$ to $g$ }}\,.
\end{align*}
}
We cast our observations in a Proposition:
\begin{proposition}
  \label{thm:UPC}
  \label{prop:USec}
  If $A$ is finite, then there is a bijective correspondence bewtween
  spaces over $P(A)$ of the form $(\Sec{\pi},\d)$ and universal
  $\Sfive^{A}$-product frames.
\end{proposition}

\subsection*{Pairwise complete spaces and lattices}
We can generalize the construction of the relational lattice $\RDA$
starting from an arbitrary space $(X,\d)$.  We say that a subset
$Z \subseteq A \cup X$ is closed if $x \in Z$ whenever
$\d(x,y) \subseteq Z \cap A$ and $y \in Z$. The set of closed subsets
of $A \cup X$ is then a Moore family.
\begin{definition}
  The lattice $\LL(X,\d)$ is the lattice of closed subsets of
  $A \cup X$.
\end{definition}
Obviously, $\LL(\AD,\d)$ is the relational lattice $\RDA$.  The
lattice $\LL(X,\d)$ can be shown to be \pperfect when $(X,\d)$ is
reduced and symmetric. Yet, for the sake of the undecidability result,
we shall only need that $\LL(X,\d)$ is an atomistic \pperfect lattice
when $A$ is finite and $(X,\d)$ is \Pc.
\begin{proposition}
  \label{cor:isolattices}
  If $A$ is finite and $(X,\d)$ is \Pc, then $\LL(X,\d)$ is isomorphic
  to the lattice $\L(\U)$ for some universal $\Sfive^{A}$-product
  frame $\U$.
\end{proposition}
\begin{proof}
  By Theorem~\ref{thm:SecPC} the space $(X,\d)$ is isomorphic to the
  space $(\Sec{\pi},\d)$, for some surjective $\pi : E \rto A$.  The
  construction $\LL$ clearly sends isomorphic spaces to isomorphic to
  isomorphic lattices. Therefore, we assume that
  $(X,\d) =(\Sec{\pi},\d)$ and prove that $\LL(X,\d) = \L(\U)$. We
  have $X = \prod_{a \in A} E_{a}$ with $E_{a} = \pi^{-1}(a)$, while
  $\d(f,g) = \set{ a \in A \mid f(a) \neq g(a)}$.
  It is easily verified that $\d(f,g) \subseteq \alpha$ if and only if
  there is an $\alpha$-path from $f$ to $g$ in the universal
  $\Sfive^{A}$ product frame $\U$ on $\prod_{a \in A} E_{a}$.
  Therefore, the two Moore families, $\LL(X,\d)$ and $\L(\U)$, are the
  same.
\end{proof}

From the above theorem and from the preliminary investigation of the
structure of the lattices $\L(\FF)$ in Section~\ref{sec:onedirection},
we can infer the following statement.

\begin{corollary}
  If $A$ is finite and $(X,\d)$ is pairwise complete, then $\LL(X,\d)$
  is an atomistic \pperfect lattice, where the set of \jirr elements
  can be identified with $A \cup X$, every element $a \in A$ is \jp,
  and \mjc{s} of $f \in X$ are of the form
  \begin{align*}
    f & \mcovered \d(f,g) \cup \set{g}\,,
  \end{align*}
  for each $g \in X$.
\end{corollary}
\begin{proof}
  The statement follows from Propositon~\ref{prop:minjoincovers} and
  from the observation that an $\alpha$-path from $f$ to $g$ is
  minimal if and only if $\alpha =\d(f,g)$.
\end{proof}
Let us remark that the above statement holds even when $A$ is not
finite or when $(X,\d)$ is not \Pc. In particular, if $A$ is infinite
and the universal $\Sfive^{A}$-product frame $\U$ has $(\Sec{\pi},\d)$
as underlying space, then the lattice $\L(\U)$ defined in
Section~\ref{sec:onedirection} and the lattice $\LL(\Sec{\pi},\d)$
defined here need not to be equal. For instance, if $\d(f,g)$ is an
infinite set, then $\d(f,g) \cup \set{g}$ is an infinite \mjc of $f$,
while we observed before that any \mjc in $\L(\U)$ is finite.

\section{Principal ideals and filters in relational lattices}
\label{sec:intervals}

The purpose of this Section is to prove the following statement.
\begin{theorem}
  \label{thm:embeddings}
  If $L$ is a finite \si atomistic lattice which has a lattice
  embedding into some relational lattice $\R(D,A)$, then there exists
  an embeddings of $L$ into some other relational lattice $\R(D,B)$
  which moreover preserves $\bot$ and $\top$.
\end{theorem}
The theorem is an immediate consequence of
Propositions~\ref{prop:embeddingtop} and~\ref{prop:embeddingtopbot}
that follows. 
These propositions mainly deal with the structure of principal ideals
and filters in a relational lattice $\R(D,A)$, namely the sublattices
of the form $\downset{Z} := \set{ W \in \R(D,A) \mid W\subseteq Z}$
and $\upset{Z} := \set{ W \in \R(D,A) \mid Z \subseteq W}$.

In the proofs of these propositions we use the isomorphism between
$P(A \sqcup \AD)$ and $P(A) \times P(\AD)$ to represent the lattice
$\R(D,A)$ as the set of pairs $(\alpha,Y)$ with $\alpha \subseteq A$
and $Y \subseteq \AD$ $\alpha$-closed.

\begin{proposition}
  \label{prop:embeddingtop}
  If $L$ is a \si lattice which has an embedding $i : L \rto \R(D,A)$,
  then there is a subset $\alpha \subseteq A$ and an embedding
  $j : L \rto \R(D,\alpha)$ that preserves $\top$.
\end{proposition}
\begin{proof}
  Suppose $i(\top) \neq (A,\AD)$, say $i(\top) = (\alpha,Y)$, with $Y$
  $\alpha$-closed. Call $M$ the ideal $\downset{(\alpha,Y)}$, so $L$
  embeds into $M$ while preserving $\top$.
  Let us study the structure of $M$. This lattice is clearly
  atomistic and \pperfect by
  Lemma~\ref{lemma:pipperfect}.
  Its set of atoms is $\alpha \cup Y$, while the
  \mjc{s} are
  of the form $f \mcovered \d(f,g) \cup \set{g}$ whenever $f,g \in Y$
  and $\d(f,g) \subseteq \alpha$.

  Notice now that if $f \in Y$, then the ball $B(f,\alpha)$ is
  contained in $Y$, since $Y$ is $\alpha$-closed.  This implies that
  $A_{f} := \alpha \cup B(f,\alpha)$ is a $D$-closed subset of
  $\Ji(M)$ and, $M_{A_{f}}$ defined in Lemma~\ref{lem:quotients}, is a
  lattice quotient of $M$. We notice that the OD-graph of $M_{A_{f}}$
  is isomorphic to the one of $\R(D,\alpha)$, so $M_{A_{f}}$ itself is
  isomorphic to $\R(D,\alpha)$.

  Since moreover
  $\bigcup_{f \in Y} \alpha \cup B(f,\alpha) = \alpha \cup Y$, then
  $\langle \pi_{A_{f}} \mid f \in Y \rangle : M \rto \prod_{f \in Y}
  M_{A_{f}}$
  is, by Lemma~\ref{lem:subdirect}, a subdirect decomposition of $M$.
  Therefore $L$ embeds into $\prod_{f \in Y} M_{A_{f}}$ and since $L$
  is \si, it embeds into some $M_{A_{f}}$ and such embedding preserves
  $\top$.  Since $M_{A_{f}}$ is isomorphic to $\R(D,\alpha)$, we
  conclude that $L$ embeds into $\R(D,\alpha)$ while preserving
  $\top$.
\end{proof}

For $B \subseteq A$, let us
define
$\psi_{A,B} : P(A) \times P(\AD) \rto P(B) \times P(\expo{B}{D})$ by
the following formula:
\begin{align*}
  \psi_{A,B}(\alpha,X) & := (\alpha \cap B, X \rrestr[B])\,.
\end{align*}
\begin{lemma}
  \label{lemma:psiAB}
  The map $\psi_{A,B}$ restricts to an \op map from $\R(D,A)$ to
  $\R(D,B)$.  Its further restriction to the filter
  $\upset \!(\complement{B},\emptyset) \subseteq \R(D,A)$ yields an
  isomorphism with $\R(D,B)$.
\end{lemma}
\begin{proof}
  We suppose that $X$ is $\alpha$-closed and argue $X\rrestr[B]$ is
  $\alpha \cap B$-closed. If $g \in X$, $f \in \BD$, and
  $\d_{\BD}(f,g\,\restr[B]) \subseteq \alpha \cap B$, then we can
  extend $f$ to $f' \in \AD$, so $f'\restr[B] = f$ and $f'(x) = g(x)$
  for all $x \in \complement{B}$. It follows that
  $\d_{\AD}(f', g) = \d_{\BD}(f,g\,\restr[B]) \subseteq \alpha \cap
  B$,
  so $f' \in X$ since $X$ is $\alpha$-closed. Then
  $f = f' \restr[B] \in X\rrestr[B]$.

  We argue similarly that if $(\beta,Y) \in \R(D,B)$, then
  $(\beta \cup \complement{B},i_{A}(Y))$ belongs to $\R(D,A)$, namely
  that $i_A(Y)$ is $\beta \cup \complement{B}$-closed when $Y$ is
  $\beta$-closed.  Let $f \in \AD$ and $g \in i_{A}(Y)$ be such that
  $\d(f,g) \subseteq \beta \cup \complement{B}$.  Now
  $\delta(f \restr[B],g\,\restr[B]) \subseteq (\beta \cup
  \complement{B}) \cap B = \beta$,
  and since $g\, \restr[B] \in Y$ and $Y$ is $\beta$-closed, we have
  $f\restr[B] \in Y$, that is $f \in i_{A}(Y)$.

  Observe moreover that
  $(\alpha \cap B,X \rrestr[B]) \subseteq (\beta,Y) $ holds if and
  only if $(\alpha,X) \subseteq (\beta \cup \complement{B},i_{A}(Y))$,
  so the two maps are adjoints to each other, in particular they are
  monotonic.

  Next
  $\psi_{A,B}(\beta \cup \complement{B},i_{A}(Y)) = ((\beta \cup
  \complement{B}) \cap B,i_{A}(Y)\rrestr[B]) = (\beta,Y)$,
  thus $\psi_{A,B}$ is surjective.
  Finally, let us argue that $\psi_{A,B}$ is injective if restricted
  to $\upset (\complement{B},\emptyset)$.  Let
  $(\alpha,X), (\alpha',X') \in \R(D,A)$ with
  $\complement{B} \subseteq \alpha \cap \alpha'$ and
  $\psi_{A,B}(\alpha,X) = \psi_{A,B}(\alpha',X')$. Then
  $\alpha \cap B = \alpha ' \cap B$,
  $\alpha = (\alpha \cap B) \cup \complement{B}$, and
  $\alpha' = (\alpha' \cap B) \cup \complement{B}$, imply
  $\alpha = \alpha'$.
  Let $f \in X$, so $f \restr[B] \in X \rrestr[B] = X' \rrestr[B]$, so
  there exists $f' \in X'$ with $f'\restr[B] = f \restr[B]$. Since
  $X'$ is $\complement{B}$ closed and
  $\d(f,f') \subseteq \complement{B}$ we have $f \in X'$. Thus we have
  $X \subseteq X'$; a similar argument yields $X' \subseteq X$, so
  $X = X'$.
\end{proof}

\begin{proposition}
  \label{prop:embeddingtopbot}
  If a finite \si atomistic lattice $L$ has a $\top$-preserving
  lattice embedding $i : L \rto \R(D,A)$, then there exists an
  embedding $j : L \rto \R(D,B)$ which preserves $\top$ and $\bot$.
\end{proposition}
\begin{proof}
  By Lemma~\ref{lemma:psiAB}, if $i(\bot) = (\complement{B},X)$ for
  some $B \subseteq A$, then $j = \psi_{A,B} \circ i : L \rto \R(D,B)$
  is an embedding which preserves $\top$ and such that
  $j(\bot) = (\emptyset,Y)$ for some $Y \subseteq \BD$. 

  Suppose now that $Y \neq\emptyset$. Let $\mu : \R(D,B) \rto L$ be
  \la to $j$, so $\mu$ is surjective and, moreover, each atom
  $a \in \Ji(L)$ has some $k \in B \cup \expo{B}{D}$ with
  $\mu(k) = a$.  Notice also that $\mu(k) = \bot$ if and only if
  $k \in Y$, for each $k \in B \cup \expo{B}{D}$.
  Let us argue that every element of $\Ji(L)$ is \jp.
  Let $a \in \Ji(L)$ and pick $k \in B \cup \BD$ such that
  $\mu(k) = a$.  If $k \in B$, then $a = \mu(k)$ is \jp, since $\mu$
  sends a \jp element either to a \jp element or to $\bot$.
  Suppose now $k = f \in \BD$ and recall that $\mu(f) = a \neq \bot$
  implies $f \not \in Y$. Pick $g \in Y$, so
  $f \leq \bigvee \d(f,g) \vee g$ and
  $a = \mu(f) \leq \bigvee \mu(\d(f,g)) \vee \mu(g) = \bv
  \mu(\d(f,g))$.
  Since $\mu$ sends \jp elements to \jp elements or to $\bot$, we see
  that $a$ has a join-cover made up of \jp elements only.
  Lemma~\ref{lemma:noprimecovers} implies then that $a$ is \jp.

  We have argued that either $Y = \emptyset$, so $j$ preserves $\bot$;
  or $Y \neq\emptyset$, in which case all the elements of $\Ji(L)$ are
  \jp and atoms. In the last case, however, $L$ is a two elements
  Boolean algebra, since $L$ is \si and distributive. Such an algebra
  can obviously be embedded into a relational lattice while preserving
  $\top$ and $\bot$.
\end{proof}

\section{From lattice embeddings to surjective \pmorphism{s}}
\label{sec:converse}

We prove in this Section the converse of Theorem~\ref{thm:direct}:
\begin{theorem}
  \label{thm:converse}
  Let $A$ be a finite set, let $\FF$ be a \fif \Sfour $A$-frame.
  If $\L(\FF)$
  embeds into a relational lattice $\R(D,B)$, then there exists a
  universal $\Sfive^{A}$-product frame $\U$ and a surjective
  \pmorphism from $\U$ to $\FF$.
\end{theorem}
To prove the Theorem, we study \bp embeddings of finite atomistic
lattices into lattices of the form $\R(D,B)$.  Let in the following
\choosedisplay{$i : L \rto \RDB$}{
\begin{align*}
  & i : L \rto \RDB
\end{align*}
} be a fixed \bp lattice embedding, with $L$ a finite atomistic
lattice. Since $L$ is finite, $i$ has a \la $\mu : \RDB \rto L$.  By
abuse of notation, we shall also use the same letter $\mu$ to denote
the restriction of this \la to the set of \cjirr elements of $\RDB$
which, we recall, is identified with  $B \cup \BD$.  It is a
general fact---and the main ingredient of Birkhoff's duality for
finite distributive lattices---that \la{s} to \bp
lattice morphism preserve \jp elements. Thus we have:
\begin{lemma}
  \label{lemma:imageofjp}
  If $b \in B$, then $\mu(b)$ is \jp.
\end{lemma}
\begin{noaiml}
  \begin{proof}
    Suppose $b \in B$ and $\mu(b) \leq \bv X$.  Then
    $b \leq i(\bv X) = \bv i(X)$, so $b \leq i(x)$ for some $x \in X$,
    since $b$ is \jp. It follows that $\mu(b) \leq x$, for some
    $x \in X$.
  \end{proof}
\end{noaiml}
It is not in general true that left adjoints send \jirr elements to \jirr
elements, and this is a main difficulty towards a proof of
Theorem~\ref{thm:converse}. Yet, the following statements hold:
\begin{lemma}
  \label{lemma:morallysurjective}
  For each $x \in \Ji(L)$ there exists $y \in B \cup \BD$ such that
  $\mu(y) = x$.
\end{lemma}
\begin{noaiml}
  \begin{proof}
    Since $i$ is an embedding, then its \la $\mu$ is
    surjective. So if $x \in \Ji(L)$, then there exists $y \in \RDB$
    with $\mu(y) = x$. Write $y = \bigvee_{i \in I} z_{i}$ with each
    $z_{i}\in B \cup \BD$. Then $x = \bigvee_{i \in I}\mu(z_{i})$, so
    $x = \mu(z_{i})$ for some $i \in I$ and such a $z_{i}$ is a
    preimage of $y$ by $\mu$ which belongs to $B \cup \BD$.
  \end{proof}
\end{noaiml}
\begin{lemma}
  \label{lemma:jreducible}
  Let $g \in \BD$ such that $\mu(g)$ is join-reducible in $L$. There
  exists $h \in \BD$ such that $\mu(h) \in \Ji(L)$ and
  $\mu(g) = \bv \mu(\d(g,h)) \vee \mu(h)$; moreover, $\mu(h)$ is \njp
  whenever $L$ is not a Boolean algebra.
  \label{lem:mnjp}
\end{lemma}
\begin{proof}
  Write $\mu(g) = \bv \alpha$ with $\alpha \subseteq \Ji(L)$ and
  $\alpha$ minimal with these two properties. We have then
  $g \leq \bv i(\alpha)$ so
  $\d(g,h) \cup \set{h} \refines i(\alpha)$ for some $h \in \BD$. We
  have then $\mu(\d(g,h)) \cup \set{\mu(h)} \refines \alpha$ and
  this relation implies that
  $\mu(\d(g,h)) \cup \set{\mu(h)} \subseteq \alpha$.  Indeed, since
  $i$ preserves the least element, $\mu(x) = \bot$ implies
  $x = \bot$.  Thus every element of
  $\mu(\d(g,h)) \cup \set{\mu(h)}$ is distinct from $\bot$ and below
  an atom in $\alpha$, so it is necessarily equal to such an
  atom. In particular, we have $\mu(h) \in \Ji(L)$.
  
  We also have
  $\bv \alpha \leq \mu(g) \leq \bv \mu(\d(g,h)) \vee \mu(h) \leq \bv
  \alpha$,
  so $\mu(g) = \bv \mu(\d(g,h)) \vee \mu(h)$. By minimality, it
  follows $\alpha = \mu(\d(g,h)) \cup \set{\mu(h)}$.
  
  Suppose that $L$ is not a Boolean algebra, so we can find an atom
  $a \in \Ji(L)$ which is \njp. Pick $f \in \BD$ such that
  $\mu(f) = a$.  Observe that every element $\d(f,g) \cup \d(g,h)$
  is \jp, so every element of $\mu(\d(f,g) \cup \d(g,h))$ is also
  \jp. If $\mu(h)$ is \jp, then we deduce
  $a \leq \bigvee \mu(\d(f,g)) \vee \bigvee \mu(\d(g,h)) \vee
  \mu(h)$,
  so the \njp $a$ has a \jc all made of \jp elements. Since $L$ is
  in the variety generated by the relational lattices, this
  contradicts Lemma~\ref{lemma:noprimecovers}.
\end{proof}
Let $A$ be the set of atoms of $L$ that are \jp.  While
$(\expo{B}{D},\d)$ is a space over $P(B)$, we need to transform
$\expo{B}{D}$ into a space over $P(A)$. To this end, we define a
$P(A)$-valued distance $\dA$ on $\BD$ by
\choosedisplay{$\dA(f,g ) := \mu(\d(f,g))$.}{
\begin{align*}
  \dA(f,g) & := \set{\mu(b) \mid b \in \d(f,g)}\,.
\end{align*}
}
Because of Lemma~\ref{lemma:imageofjp}, we have
$\dA(f,g) \subseteq A$.
\begin{proposition}
  $(\BD,\dA)$ is a pairwise complete ultrametric space over $P(A)$.
\end{proposition}
\begin{proof}
  $\dA$ satisfies the properties defining a distance (including being
  reduced and symmetric), mainly because the direct image of any
  function (here of $\mu$) preserves unions.

  For \PC, observe that if
  $\dA(f,g) \subseteq \alpha_{0} \cup \alpha_{1}$, then
  $\d(f,g) \subseteq \beta_{0} \cup \beta_{1}$, where
  $\beta_{i} := \set{b \in B \mid \mu(b) \in \alpha_{i}}$, $i = 0,1$.
  Taking $h$ such that $\delta(f,h) \subseteq \beta_{0}$ and
  $\d(h,g) \subseteq \beta_{1}$, we obtain
  $\dA(f,h) \subseteq \alpha_{0}$ and $\dA(h,g) \subseteq \alpha_{1}$.
\end{proof}

We define next $v: \expo{B}{D} \rto P(A)$ by letting
\choosedisplay{$v(f) := \set{ a \in A \mid a \leq \mu(f)}$.}{
\begin{align*}
  v(f) & := \set{ a \in A \mid a \leq \mu(f)}\,.
\end{align*}
}
\begin{lemma}
  $v : \BD \rto P(A)$ is a module on $(\BD,\dA)$. 
\begin{noaiml}
    That is, the relation
    \choosedisplay{$v(f) \subseteq \dA(f,g) \cup v(g)$}{
      \begin{align*}
        v(f) & \subseteq \dA(f,g) \cup v(g)\,.
      \end{align*}
    } holds.
\end{noaiml}
\end{lemma}
\begin{proof}
  Suppose that $a \in v(f)$ and $a \not\in \dA(f,g)$. This means that
  $a \leq \mu(f)$ but $b \not\in \d(f,g)$ whenever $\mu(b) = a$.
  Recall that if $b \in B$, then $b$ is \jp, whence $\mu(b)$ is \jp as
  well. Thus if $a \in A$ and $a \leq \mu(b)$, then $a = \mu(b)$,
  since we are assuming that $L$ is atomistic.
  Since
  $a \leq \mu(f) \leq \bigvee_{b \in \delta(f,g)} \mu(b)\vee \mu(g)$,
  $a$ is \jp, $a \leq \mu(b)$ implies $a = \mu(b)$, we necessarily
  have $a \leq \mu(g)$, so $a \in v(g)$.
\end{proof}

\begin{lemma}
  \label{lemma:VJLmB}
  The map $v : \BD \rto P(A)$ is a module on $(\BD,\dA)$.  Moreover
  $v(f) = \emptyset$ if and only if $\mu(f) \in \Ji(L)\setminus A$.
\end{lemma}
\begin{proof}
  Suppose that $\mu(f) \in \Ji(L) \setminus A$.  If
  $v(f) \neq \emptyset$, then let $a \in A$ with $a \leq \mu(f)$.
  Since we are assuming that $\mu(f)$ is \jirr and that $L$ is
  atomistic, we deduce $\mu(f) = a \in A$, \contr.

  Conversely, suppose that $v(f) = \emptyset$. This immediately gives
  $\mu(f) \not \in A$. \Bcontr, suppose now that $\mu(f)$ is
  reducible, so use Lemma~\ref{lem:mnjp} to find $h \in \BD$ such that
  $\mu(h) \in \Ji(L)$ and $\mu(f) = \bv \mu(\d(f,h)) \vee
  \mu(h)$.
  Since $\mu(h)$ is \jirr, then $h \neq f$ and
  $\d(f,h) \neq \emptyset$.  Pick $b \in \d(f,h)$, then
  $\mu(b) \in A$ and $\mu(b) \leq \mu(f)$.  This gives
  $\mu(b) \in v(f)$, so $v(f) \neq \emptyset$, \contr.
\end{proof}

Using Lemmas~\ref{lemma:kernmodulepairwisecomplete}
and~\ref{lemma:VJLmB}, we derive:
\begin{corollary}
  \label{cor:Fz}
  The subspace
  \choosedisplay{$\Fz := \set{ f \in \BD \mid \mu(f) \in \Ji(L) \setminus A}$}{
    \begin{align}
      \label{def:Fz}
      \Fz & := \set{ f \in \BD \mid \mu(f) \in \Ji(L) \setminus A}
    \end{align}
  }
  of $\BD$ is
  \Pc.
\end{corollary}
\begin{proof}
  By Lemma~\ref{lemma:VJLmB}, $f \in \Fz$ if and only if
  $v(f) = \emptyset$. Since $v$ is a module, the set
  $\set{ f \in \BD\mid v(f) = \emptyset}$ is, by
  Lemma~\ref{lemma:kernmodulepairwisecomplete}, a \Pc metric space
  over $P(A)$.
\end{proof}
The following Proposition, which ends the study of \bp lattice
embeddings into relational lattices, shows that modulo the shift of
the codomain to the lattice of a universal product frame, such a
lattice embedding can always be normalized, meaning that \jirr
elements are sent to \jirr elements by the left adjoint.
\begin{proposition}
  \label{prop:regularembedding}
  Let $L$ be a finite atomistic lattice and let $A$ be the set of its
  \jp elements.  If $L$ is not a Boolean algebra and
  $i : L \rto \R(D,B)$ is a \bp lattice embedding, then there
  exists a \Pc ultrametric space $(\Fz,\d)$ over $P(A)$ and a \bp
  lattice embedding $j : L \rto \LL(\Fz,\dA)$ whose \la $\nu$
  satisfies the following condition: for each $k \in A \cup \Fz$, if
  $k \in A$ then $\nu(k) = k$ and, otherwise,
  $\nu(k) \in \Ji(L) \setminus A$.
\end{proposition}
\begin{proof}
  Let $(\Fz,\dA)$ be the \Pc space over $P(A)$ as defined in
  Corollary~\ref{cor:Fz}.
  By the definition of $\Fz$, the
  restriction of $\mu$ to $\Fz$ takes values in $\Ji(L)\setminus A$.
  Therefore we can define $\nu : A \cup \Fz \rto \Ji(L)$ 
\begin{aiml}
  by saying that $\nu(k) := k$ if $k \in A$, and $\nu(k) := \mu(k)$, otherwise. 
\end{aiml}
\begin{noaiml}
  as follows:
  \begin{align*}
    \nu(k) & :=
    \begin{cases}
      k\,, &  k \in A, \\
      \mu(k)\,, & k \in \Fz.
    \end{cases}
  \end{align*}
\end{noaiml}
  We notice next that $\nu$ is surjective. If
  $x \in \Ji(L)\setminus A$, then by
  Lemma~\ref{lemma:morallysurjective}, there is $y \in B \cup \BD$
  such that $\mu(y) = x$. By Lemma~\ref{lemma:imageofjp},
  $y \not\in B$, so $y \in \BD$. Since
  $\mu(y) = x \in \Ji(L) \setminus A$, then $y$ belongs to $\Fz$.

  Let $j : L \rto \LL(\Fz,\dA)$ be the function defined by
  \choosedisplay{$j(l) := \set{ x \in A \cup \Fz \mid \nu(x) \leq
      l}$.
  }{
  \begin{align*}
    j(l) & := \set{ x \in A \cup \Fz \mid \nu(x) \leq l}\,.
  \end{align*}
  }
   Let us argue, in the order, that
\begin{aiml}
     (1) for each $l \in L$, $j(l)$ is a closed subset of
       $A \cup \Fz$, (2) $j$ is injective,
     (4) $j$ preserves meets and (4) it preserves joins.
\end{aiml}
\begin{noaiml}
     \begin{enumerate}[(1)]
     \item for each $l \in L$, $j(l)$ is a closed subset of
       $A \cup \Fz$,
     \item $j$ is injective,
     \item $j$ preserves meets and (4) it preserves joins.
     \end{enumerate}
\end{noaiml}

  (1). Let $f,g \in \Fz$ and suppose that 
  $\dA(f,g) \cup \set{g} \subseteq j(l)$. This condition means
  that $\nu(\dA(f,g)) = \dA(f,g) = \mu(\d(f,g)) \refines \set{l}$ and
  $\mu(g) \leq l$; it follows that
  $\nu(f) = \mu(f) \leq \bv \mu(\d(f,g)) \vee \mu(g) \leq l$, so
  $f \in j(l)$.
  
  (2). We have $j(l_{0}) = j(l_{1})$ if and only if, for all
  $x \in A \cup \Fz$, the condition $\nu(x) \leq l_{0}$ is equivalent to
  $\nu(x) \leq l_{1}$. As $\nu$ is surjective, this means that
  $l_{0}$ and $l_{1}$ have the same atoms below them, thus that they
  are equal.
  
  (3).  It is easily verified that
  \choosedisplay{$j(\top) = A \cup \Fz$ and $j(l_{0} \land
    l_{1}) = j(l_{0}) \cap j(l_{1})$.
  }{\begin{align*}
    j(\top) & = A \cup \Fz \,, \quad\text{and}\quad j(l_{0} \land
    l_{1}) = j(l_{0}) \cap j(l_{1})\,.
  \end{align*}}
  In particular, $j$ is \op.
  
  (4). Since $j$ is \op, we only need to show that
  $j(l_{0} \vee l_{1}) \leq j(l_{0}) \vee j(l_{1})$.
  To this end, we suppose that $x \in A \cup \Fz$ is such that
  $x \in j(l_{0} \vee l_{1})$, so $\nu(x) \leq l_{0} \vee l_{1}$.  If
  $x \in A$, then $\nu(x) = x \leq l_{0} \vee l_{1}$, and since $x$ is
  \jp, this gives $\nu(x) = x \leq l_{i}$ for some $i \in \set{0,1}$.
  This immediately yields
  $x \in j(l_{i}) \leq j(l_{0}) \vee j(l_{1})$.

  Suppose now that $x = f \in \Fz$ so
  $\mu(f)= \nu(f) \leq l_{0} \vee l_{1}$.  We have, therefore,
  $f \leq i(l_{0}) \vee i(l_{1})$, so
  $f \mcovered \d(f,g) \cup \set{g} \refines \set{i(l_{0}),i(l_{1})}$
  for some $g \in \BD$.  We can use now Lemma~\ref{lem:mnjp} to pick
  $h \in \BD$ with $\mu(h) \in \Ji(L) \setminus A$ (so $h \in \Fz$)
  and $\mu(g) = \bv \mu(\d(g,h)) \vee \mu(h)$.

  We have then that
  $\mu(\d(f,h)) \cup \set{\mu(h)} \subseteq \mu(\d(f,g) \cup \d(g,h))
  \cup \set{\mu(h)} \refines \set{l_{0},l_{1}}$.
  This relation yields
  \choosedisplay{$\nu(\dA(f,h)) \cup \set{\nu(h)} \refines
    \set{l_{0},l_{1}}$}{
  \begin{align*}
    \nu(\dA(f,h)) \cup \set{\nu(h)} & \refines \set{l_{0},l_{1}}
  \end{align*}
  }
  or, said otherwise, 
  \choosedisplay{$\dA(f,h) \cup \set{h} \refines
    \set{j(l_{0}),j(l_{1})}$.}
  {
  \begin{align*}
    \dA(f,h) \cup \set{h} & \refines \set{j(l_{0}),j(l_{1})}\,.
  \end{align*}
  }
  This implies that $f \in j(l_{0}) \vee j(l_{1})$.

  Let us argue that $j$ preserves the least element. If $x \in j(\bot)$, then $\nu(x) \leq \bot$.  We cannot have
  $x \in B$, so $x = f \in \Fz$. Then $\mu(f) \leq \bot$ and
  $f \in i(\bot)$, contradicting the assumption that $i$ preserves
  bounds.

  Finally, let us observe that the \la of $j$ agrees, on
  \jirr elements, with $\nu$. Indeed, for each $x \in A \cup \Fz$, we
  have $\nu(x) \leq y$ iff $x \in j(y)$, iff $x \leq j(y)$, where we
  identify, as usual, a singleton with its only element.
\end{proof}

\begin{aiml}
  For lack of space
  let us state the following Theorem without  a
  proof.
  \begin{theorem}
    \label{thm:embeddings}
    If $L$ is a finite \si atomistic lattice which has a lattice
    embedding into some relational lattice $\R(D,A)$, then there exists
    a \bp embeddings of $L$ into some other relational lattice $\R(D,B)$.
  \end{theorem}
\end{aiml}

We conclude next the proof of the main result of this Section,
Theorem~\ref{thm:converse}.
\begin{myproof}{Proof of Theorem~\ref{thm:converse}}
  Since $\FF$ is \rooted and full, $\L(\FF)$ is a finite atomistic \si
  lattice by Proposition~\ref{prop:LFFsi}. Therefore, if
  $i : \L(\FF) \rto \R(D,B)$ is a lattice embedding, then we can
  assume, using Theorem~\ref{thm:embeddings}, that $i$ preserves the
  bounds. Also, if $\L(\FF)$ is a Boolean algebra, then it is the two
  elements Boolean algebra, since we are assuming that $\L(\FF)$ is
  \si. But then, $\FF$ is a singleton, and the statement of the
  Theorem trivially holds in this case.

  We can therefore assume that $\L(\FF)$ is not a Boolean algebra. Let
  us recall that $A$ is the set of \jp elements of $\L(\FF)$, see
  Proposition~\ref{prop:minjoincovers}. According to in
  Proposition~\ref{prop:regularembedding}, let $(\Fz,\dA)$ be the \Pc
  space over $P(A)$ and let $j : \L(\FF) \rto \LL(\Fz,\dA)$ be the
  lattice morphism with the properties stated there; let $\nu$ be the
  \la to $j$.  Using Corollary~\ref{cor:isolattices}, we can also
  assume that $\LL(\Fz,\dA) = \L(\U)$ for some universal
  $\Sfive^{A}$-product frame $\U$.
  
  To avoid confusions, we depart from now on from the convention of
  identifying singletons with their elements.
  We define $\psi : X_{\U} \rto X_{\FF}$ by saying that $\psi(x) = y$
  when $\nu(\set{x}) = \set{y}$. This is well defined since in
  $\L(\U)$ (respectively $\L(\FF)$) the \njp \jirr-elements are the
  singletons $\set{x}$ with $x \in \X[\U]$ (resp.  $x \in \X$);
  moreover, we have $X_{\U} = \Fz$ and each singleton $\set{x}$ with
  $x \in \Fz$ is sent by $\nu$ to a singleton
  $\set{y} \in \Ji(\L(\FF)) \setminus \set{\set{a} \mid a \in A} =
  \set{ \set{x} \mid x \in X_{\FF} }$.
  The function $\psi$ is surjective since every \njp atom $\set{x}$ in
  $\L(\FF)$ has a preimage by $\nu$ an atom $\set{y}$ and such a
  preimage cannot be \jp, so $y \in X_{\U}$.

  We are left to argue that $\psi$ is a \pmorphism. To this end, let
  us remark that, for each $a \in A$ and $x,y \in \X$ (or
  $x,y \in X_{\U}$), the relation $x R_{a} y$ holds exactly when there
  is an $\set{a}$-path from $x$ to $y$, i.e. when
  $\set{x} \subseteq \closure{\set{a,y}} = \set{a} \vee \set{y}$ (we
  need here that $\FF$ and $\U$ are \Sfour frames).

  Thus, let $x,y \in X_{\U}$ be such that $x R_{a} y$.  Then
  $\set{x} \subseteq \set{a} \vee \set{y}$ and
  $\nu(\set{x}) \subseteq \nu(\set{a}) \vee \nu(\set{y}) = \set{a}
  \vee \nu(\set{y})$.  We have therefore $\psi(x) R_{a} \psi(y)$.
  Conversely, let $x\in X_{\U}$ and $z \in \X$ be such that
  $\psi(x) R_{a} z$.  We have therefore
  $\nu(\set{x}) \subseteq \set{a} \vee \set{z}$, whence, by
  adjointness,
  \begin{align*}
    \set{x} \subseteq j(\set{a} \vee \set{z}) & = j(\set{a}) \vee
    j(\set{z})\\ 
    &= \set{a} \vee \set{y \mid
      \nu(\set{y}) = \set{z}} \\
    & = \closure{\set{a} \cup \set{ y \mid \nu(\set{y}) = \set{z}}}\,.
  \end{align*}
  But this means that 
  there is some $y \in X_{\U}$ with $\psi(y) = z$ and a $\set{a}$-path
  from $x$ to $y$. But then, we also have $x R_{a} y$.
\end{myproof}

\bibliographystyle{abbrv}
\bibliography{biblio}

\end{document}